\pdfoutput=1
\documentclass[a4paper,11pt]{article}
\usepackage{jcappub} 
\usepackage{lineno}
\usepackage{caption}
\usepackage{subcaption}
\usepackage{soul}
\usepackage[numbers]{natbib}

\title{ Modification of the Dipole in Arrival Directions of Ultra-high-energy Cosmic Rays due to the  Galactic Magnetic Field}

\author[1]{A. Bakalová,}
\author[1]{J. Vícha,}
\author[1]{P. Trávníček}
\affiliation[1]{FZU - Institute of Physics of the Czech Academy of Sciences,\\
Na Slovance 1999/2, 182 21 Prague 8, Czech Republic}

\emailAdd{bakalova@fzu.cz}

\abstract{The direction and magnitude of the dipole anisotropy of ultra-high-energy cosmic rays with energies above 8 EeV observed by the Pierre Auger Observatory indicate their extragalactic origin. The observed dipole on Earth does not necessarily need to correspond to the anisotropy of the extragalactic cosmic-ray flux due to the effects of propagation in the Galactic magnetic field. We estimate the size of these effects via numerical simulations using the CRPropa 3 package.  The Jansson--Farrar and Terral--Ferrière models of the Galactic magnetic field are used to propagate particles within the Galaxy. We identify allowed directions and amplitudes of the dipole outside the Galaxy that are compatible with the measured features of the dipole on Earth for various mass composition scenarios at the 68\% and 95\% confidence level.}

\begin{document}
\maketitle
\flushbottom

\section{Introduction}
\label{sec:intro}

Our knowledge about ultra-high-energy cosmic rays (above $10^{18}$~eV, UHECRs) grew significantly in the last two decades. Large-area observatories, like the Pierre Auger Observatory \cite{PAO} in the Southern hemisphere and Telescope Array \cite{TA} in the Northern hemisphere measure cosmic rays at ultra-high energies with an unprecedented precision, opening possibilities to study the properties of UHECRs in more detail than ever before. 

Even though the existence of UHECRs was confirmed more than 50 years ago, the sources of these particles and the mechanisms of their acceleration are still unknown and are a subject of abounding studies. Currently, there are several pieces of strong evidence suggesting an extragalactic origin of these particles. 
Searches for correlation of arrival directions of UHECRs with astrophysical sources such as active galactic nuclei or starbursts galaxies are being performed \cite{AnisotropyPAO2022, AnisotropyTA2018} with results suggesting an anisotropic distribution of these particles above $\approx 40$~EeV. An argument supporting the extragalactic origin of UHECRs is the fact that no significant anisotropies have been observed at higher energies around the Galactic plane or the Galactic center \cite{Auger2015AD}.
Nonetheless, the true sources are still unknown. The hypothesis about an extragalactic origin of UHECRs is strongly supported by a measurement of dipole anisotropy in the arrival directions of cosmic rays with energies above 8~EeV by the Pierre Auger Observatory \cite{AugerDipole2017, AugerDipole2018}. These measurements show that the direction of the dipole anisotropy points $\approx 125^{\circ}$ away from the Galactic center with uncertainty of about $15^{\circ}$,  which is much more than deflections expected by current models of the Galactic magnetic field. The amplitude of the measured dipole in the arrival directions of cosmic rays above 8~EeV is $\approx6.5~\%$ with more than $6\sigma$ significance. At lower energies, in the energy bin from $4$~EeV up to $8$~EeV, the reconstructed dipole has an amplitude of $\approx2.5~\%$ with significance smaller than $3\sigma$ \cite{AugerDipole2018}.

Trajectories of UHECRs are influenced by magnetic fields in the Universe on their long journeys from their sources to Earth. Deflections of cosmic rays in magnetic fields depend on the charge $Z$ and energy $E$ of the particle and the specifics of the magnetic fields they are crossing. The Galactic magnetic field (GMF) is not well understood. However, multiple models of the field have been proposed, for example see \cite{JF12, JF12b, TF17, PT11}, based on available observations of synchrotron radiation or Faraday rotation measurements. 
As the GMF is considered to be much stronger than the extragalactic magnetic fields in the Universe \cite{Widrow2002}, it plays an important role in the particle deflection from its source direction even though the propagation distance is much smaller within the Galaxy than in the extragalactic space.
The presence of the GMF is not only causing deflections at the single-particle level but, as a consequence, it can have a significant influence on the observed large-scale anisotropies as well.  The GMF can also change amplitudes of anisotropies and smear them as the field tends to isotropise particle trajectories, especially at lower rigidities ($R=E/Z$). These effects were shown for example in \cite{EichmanmDipole}.

In this work, we investigate the influence of the GMF on the large-scale dipole anisotropy in arrival directions of UHECRs above $8$ EeV employing numerical simulations of cosmic-ray propagation in the GMF using the CRPropa 3 software \cite{CRPropa3}. We assume a dipole anisotropy in the arrival directions of extragalactic cosmic rays at the edge of the Galaxy. We investigate the shift of the dipole direction and the change of the dipole amplitude due to the deflections of cosmic rays in the GMF during their propagation. Specifically, we are searching for dipoles in distributions of extragalactic cosmic rays that are, after propagation in the GMF, compatible with the dipole measured on Earth by the Pierre Auger Observatory. We study two models of the GMF and various mass-composition scenarios of the overall cosmic ray flux. Visualisation of the GMF models can be found in Appendix~\ref{A:GMF}.

The paper is structured as follows: We describe the used simulation settings of the cosmic-ray propagation in the GMF and the method of dipole reconstruction in Section~\ref{sec:sims}. The influence of the GMF on the dipole properties in general is described in Section~\ref{sec:GMFinfluence}. In Section~\ref{sec:results}, we derive the sky regions for allowed directions of the extragalactic dipole that are compatible with the measured dipole properties on Earth. We discuss and summarise our findings in Sections~\ref{sec:dis} and \ref{sec:conc}, respectively.

\section{Simulations and Method }
\label{sec:sims}

\subsection{Simulating Cosmic-ray Propagation}

For the study of the influence of the GMF on the dipole anisotropy in arrival directions, we simulate cosmic-ray propagation within the Galaxy using CRPropa~3 (version 3.1.7). We use backtracking of antiparticles from an isotropically emitting source positioned in coordinates $(x,y,z)=(-8.5,0,0)$~kpc, corresponding to the location of the Solar system in the Galaxy. Particles are collected at the edge of the Galaxy in a distance of 20 kpc from the Galactic center in coordinates $(x,y,z)=(0,0,0)$~kpc. The energy losses on photon backgrounds are neglected since the energy-loss lengths are much larger than the distance traveled by a particle in Galaxy \cite{Batista_2015}, even when accounting for deflections in the GMF. Therefore, according to the time-charge invariance, the propagation of antiparticles from the observer to the edge of the Galaxy is compatible with tracing particles coming from the opposite direction \cite{Kaapa, CRPropa3}. The influence of the extragalactic field is not taken into account, see Appendix~\ref{B:EGMF}.

Propagation of four types of particles is simulated; protons ($^1$H), helium ($^4$He), nitrogen ($^{14}$N), and iron ($^{56}$Fe) nuclei. Each element is simulated separately, with a power-law energy spectrum with spectral index $\gamma=3$, which is close to the observed spectral index of cosmic rays above $8\,\rm{EeV}$ \cite{PAOenergyspectrum, TAenergyspectrum}, see also Appendix \ref{D:SI}. Particles are simulated separately in two energy ranges, from $8$~EeV up to $100$~EeV, and for lower energies from 4~EeV up to 8~EeV. We simulate 250,000 particles for each element and each energy range. The particle simulation is aborted and not considered if it traveled more than $500\,\rm{kpc}$. However, in our simulations no particles were rejected by this criterion, therefore also the spectral index after the propagation remains unchanged as the energy losses are being neglected and no particles are lost.

For this study, we propagate particles in multiple variations of two models of the GMF. The Jansson--Farrar 2012 model of the GMF \cite{JF12, JF12b}, that consists of a regular component and random large scale (striated) and random small scale (turbulent) components is used. We use the updated values of the parameters of the model as suggested by the Planck collaboration \cite{JF12Planck} and we denote the model as JF12Planck. When compared to the original model \cite{JF12, JF12b}, the most significant change in the model with the updated parameters is the decrease of the strength of the random component and changes of the field strength in the spiral arms \cite{CRPropa32, JF12Planck}. We performed simulations of particle propagation in multiple realisations of the random components of the field with different seeds for three different coherence lengths $L_c=30$~pc, $L_c=60$~pc and $L_c=100$~pc \cite{farrar2014}. The model was used in superposition with a model of the magnetic field in the central mass zone of the Galaxy \cite{CMZ, CRPropa32} as the model does not implement any field in this central region. Visualization of the field strength is shown in Figure~\ref{fig:GMFJF12Planck} in Appendix~\ref{A:GMF}. The following results will be shown as a combination of all realisations of the JF12Planck simulations if not stated otherwise. 

The second model used in this work is the Terral--Ferrière model of the GMF (TF17) \cite{TF17}. There are three models of the galactic disk proposed in \cite{TF17} together with two models of the galactic halo, a bisymmetric and an antisymmetric one. The disk and halo models can be combined together, creating 6 different models of the GMF. As the bisymetric model of the halo field fits the data slightly better than the axissymmetric halo field, we use the three proposed models of the GMF with bisymmetric halo that are denoted as Ad1C1, Bd1C1 and Dd1C1. Visualisations of different options of the TF17 model of the GMF are shown in Figures~\ref{fig:GMFAd1},~\ref{fig:GMFBd1},~\ref{fig:GMFDd1} in Appendix~\ref{A:GMF}. Note that the field has quite large local values of the magnetic field strength compared to the JF12Planck model, especially in the case of the Ad1C1 and Bd1C1 options.

\subsection{Dipole Reconstruction from Simulated Data}

Information about each particle at the start and end point of the simulation is kept including their energy, unit momentum vector, and Cartesian coordinates. The isotropic flux of the simulated particles is reweighted according to the travel direction of cosmic rays at the edge of the Galaxy in order to obtain a dipole distribution of the particles arriving into Galaxy. A weight $w$ is assigned to every particle according to 

\begin{equation}
w=A_0\,\cos\delta + 1,   
\end{equation}
where $A_0$ is the extragalactic amplitude of the dipole and $\delta$ is the angular distance of the direction of the initial momentum of the particle and that of the imposed extragalactic dipole direction. The amplitude is expressed as a percentage of the relative excess with respect to the mean flux. We investigate the initial amplitude of the dipole $A_0$ in the range from $6.5$\% up to $20$\% in discrete steps of 2\% (the first step is 1.5 from amplitude 6.5\% to 8\%).  The amplitude $A_0$ of the extragalactic dipole should be equal or higher than the one observed on Earth due to the effect of isotropisation of the cosmic-ray flux in the GMF.  These amplitudes are imposed through the weights into the particle flux in all possible extragalactic directions of the dipole with a step of $1^{\circ}$ in longitude and $1^{\circ}$ in latitude.

Characteristics of the three-dimensional dipole on the observer level are reconstructed using a procedure from \cite{3Ddipole}. To calculate the direction {$\textbf{D}_{\rm{obs}}$} and amplitude $A_{\rm{obs}}$ of the dipole on the observer we need to estimate discrete version of the zeroth and first moments of the flux on the observer as

\begin{equation}
    S_0=\sum_{k}\frac{1}{w_k}
\,\,\,\,\,\,\,\,\rm{and}\,\,\,\,\,\,\,\, 
    \textbf{S}=\sum_{k}\frac{\textbf{u}_k}{w_k},
\end{equation}
where the sums go over all particles $k$ reaching the observer with weight $w_k$. {$\textbf{u}_k$} is their arrival direction to the observer. Using these we can calculate the amplitude $A_{\rm{obs}}$ and direction ${\textbf{D}_{\rm{obs}}}$ of the dipole on the observer as
\begin{equation}
\label{eq:reco3d}
    A_{\rm{obs}}=3\frac{\left\| \textbf{S} \right\|}{S_0} \,\,\,\,\,\,\,\,\rm{and}\,\,\,\,\,\,\,\, {\textbf{D}_{\rm{obs}}}=\frac{\textbf{S}}{\left\| \textbf{S} \right\|}.
\end{equation}

We are identifying solutions with extragalactic amplitudes of the dipole $A_0$ and extragalactic directions of the dipole $(l_0, b_0)$, where $l_0$ is the galactic longitude and $b_0$  galactic latitude, that are compatible with the measurement by the Pierre Auger Observatory \cite{AugerDipole2018} within $1\sigma$ and $2\sigma$ in both direction and amplitude, i.e. an amplitude of $6.5^{+1.3}_{-0.9}$\% and direction with right ascension $\alpha_{\rm{d}}=(100\pm10)^{\circ} $ and declination $\delta_{\rm{d}}=(-24^{+12}_{-13})^{\circ}$.


\section{Influence of the GMF on the Dipole Properties}
\label{sec:GMFinfluence}
\subsection{Dipole Amplitude}
As a result of diffusive propagation of cosmic rays in the GMF, the amplitude of the dipole on the observer $A_{\rm{obs}}$ (Earth) is lower than the dipole amplitude outside of the Galaxy ($A_0$). The level of the decrease of the amplitude depends strongly on the particle rigidity. Particles with lower rigidities are more deflected in the GMF during their travel and the final flux tends to isotropise. The predictions of the decrease of amplitude differ for individual models of GMF. 

\begin{figure}[htp]
     \centering
         \includegraphics[width=0.6\textwidth]{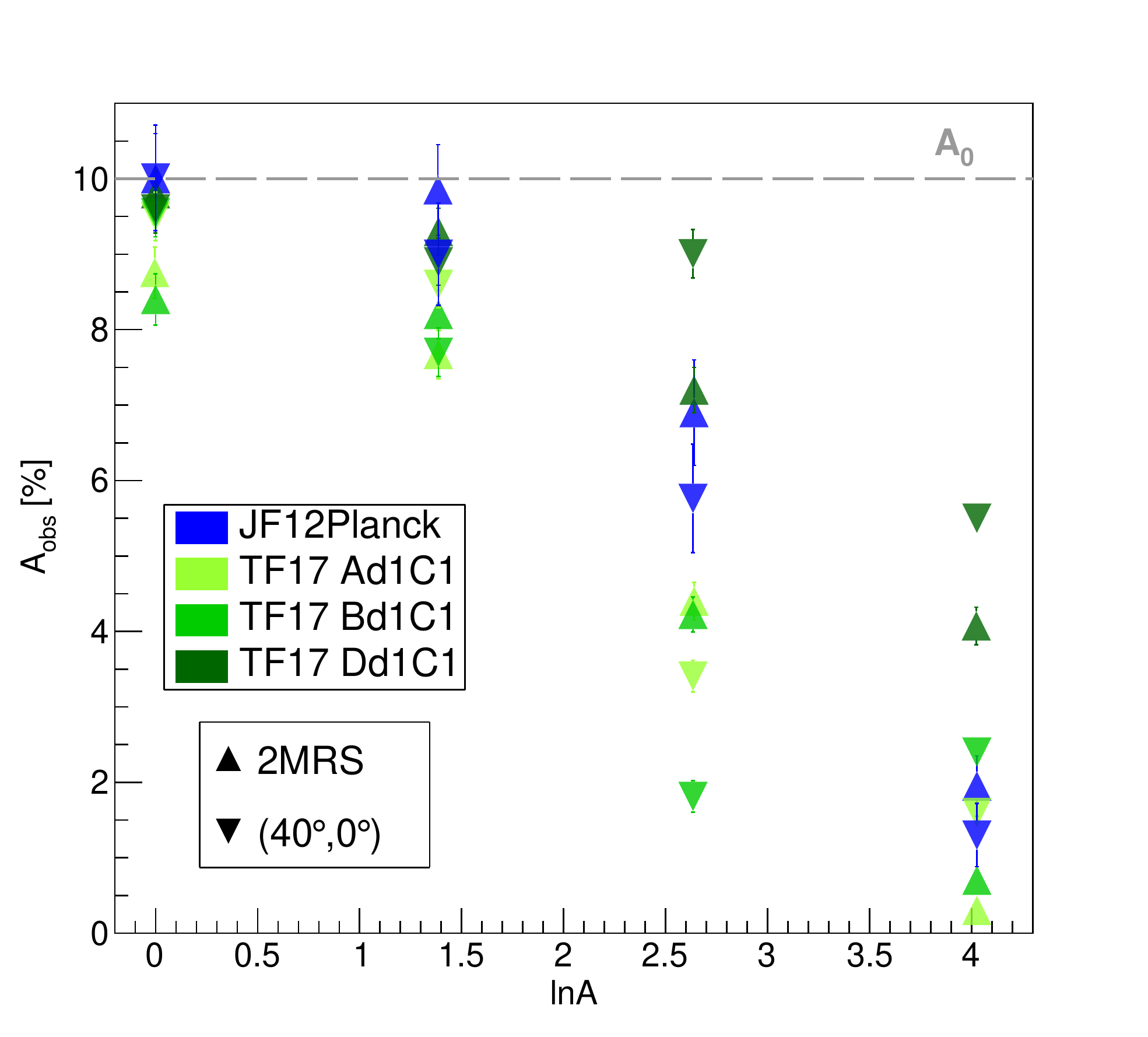}
        \caption{Reconstructed dipole amplitude on the observer for different types of particles of energies above 8~EeV with mass $A$ propagated in the TF17 (green) and the JF12Planck (blue) models of the GMF. The imposed dipole amplitude ($A_{0}$) is 10\% for two extragalactic directions of the dipole; The example direction was chosen to the 2MRS dipole in coordinates $(l_0, b_0)=(251^{\circ}, 37^{\circ})$ and direction $(l_0, b_0)=(40^{\circ},0^{\circ})$. The error bars represent $1\sigma$ certainty levels.}
        \label{fig:amplitudeChange}
\end{figure}

Not all the imposed extragalactic directions of the dipole lead to the same decrease of the amplitude within given GMF model. The change of the amplitude for different cosmic-ray species and two different directions of the dipole are depicted in Figure~\ref{fig:amplitudeChange} for cosmic rays propagated in the JF12Planck and the three options of the TF17 model of the GMF. The imposed extragalactic amplitude is $A_{0}=10$\%. One of the imposed directions corresponds to the 2MRS dipole direction \cite{2MRS} and the other direction, $(l_0, b_0)=(40^{\circ}, 0^{\circ})$, is chosen randomly in order to demonstrate the different behavior of the amplitude suppression and does not correspond to any significant astrophysical system. In case of the JF12Planck, the amplitude on the observer is calculated as the mean value from different realisations of the field. We see different predictions of the amplitude decrease at the observer for the different options of the TF17 filed, most notably for nitrogen nuclei scenario and dipole in the direction $(l_0, b_0)=(40^{\circ}, 0^{\circ})$. While  the Dd1C1 model gives an amplitude on the observer of 9.1\%, the Bd1C1 model shows a suppression of the amplitude down to lower  than 2\%.

\begin{figure}
     \centering
     \begin{subfigure}{0.49\textwidth}
         \centering
          \includegraphics[width=1.0\textwidth]{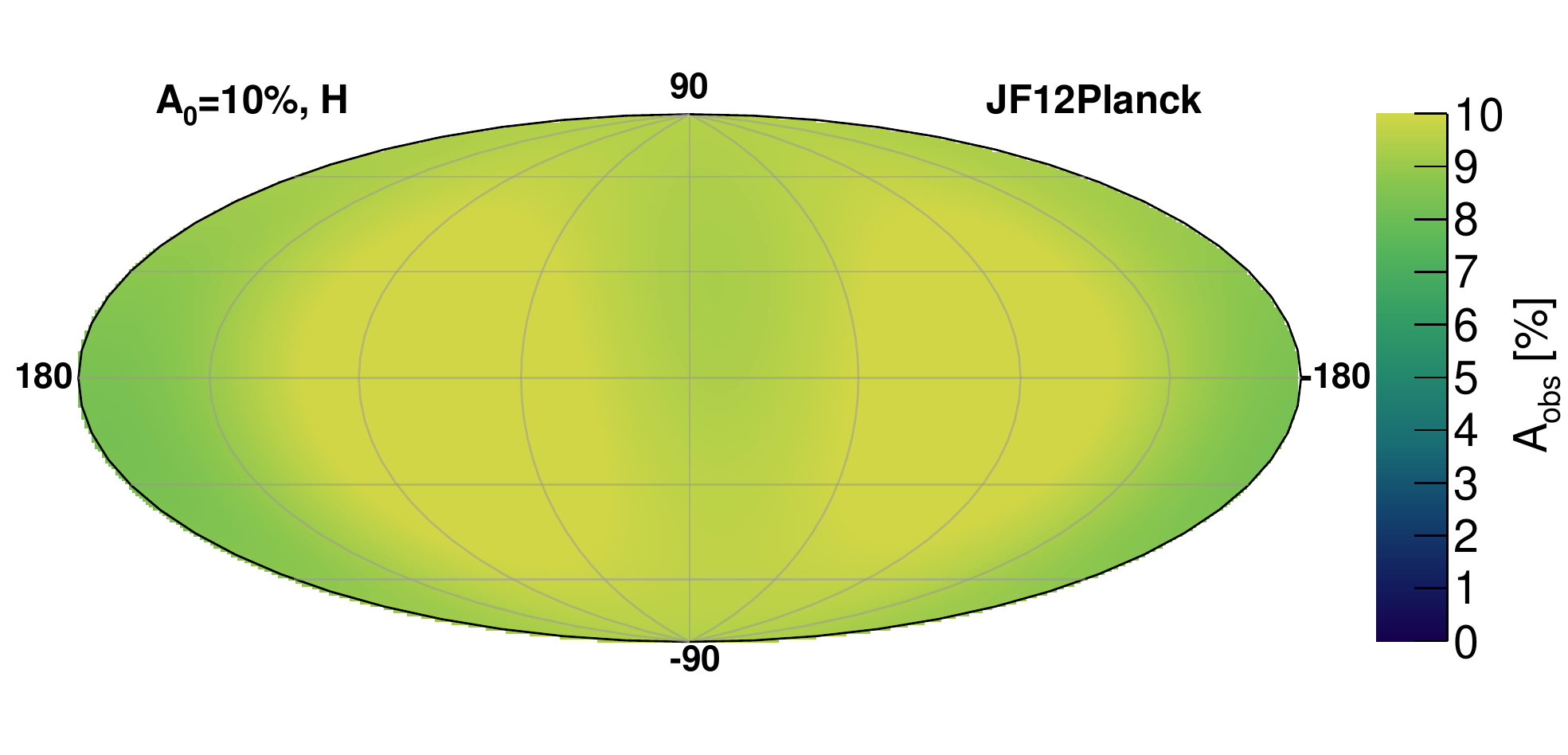}
         
         \label{fig:amplitudeH_JF12}
     \end{subfigure}
     \hfill
     \begin{subfigure}{0.49\textwidth}
         \centering
        \includegraphics[width=1.0\textwidth]{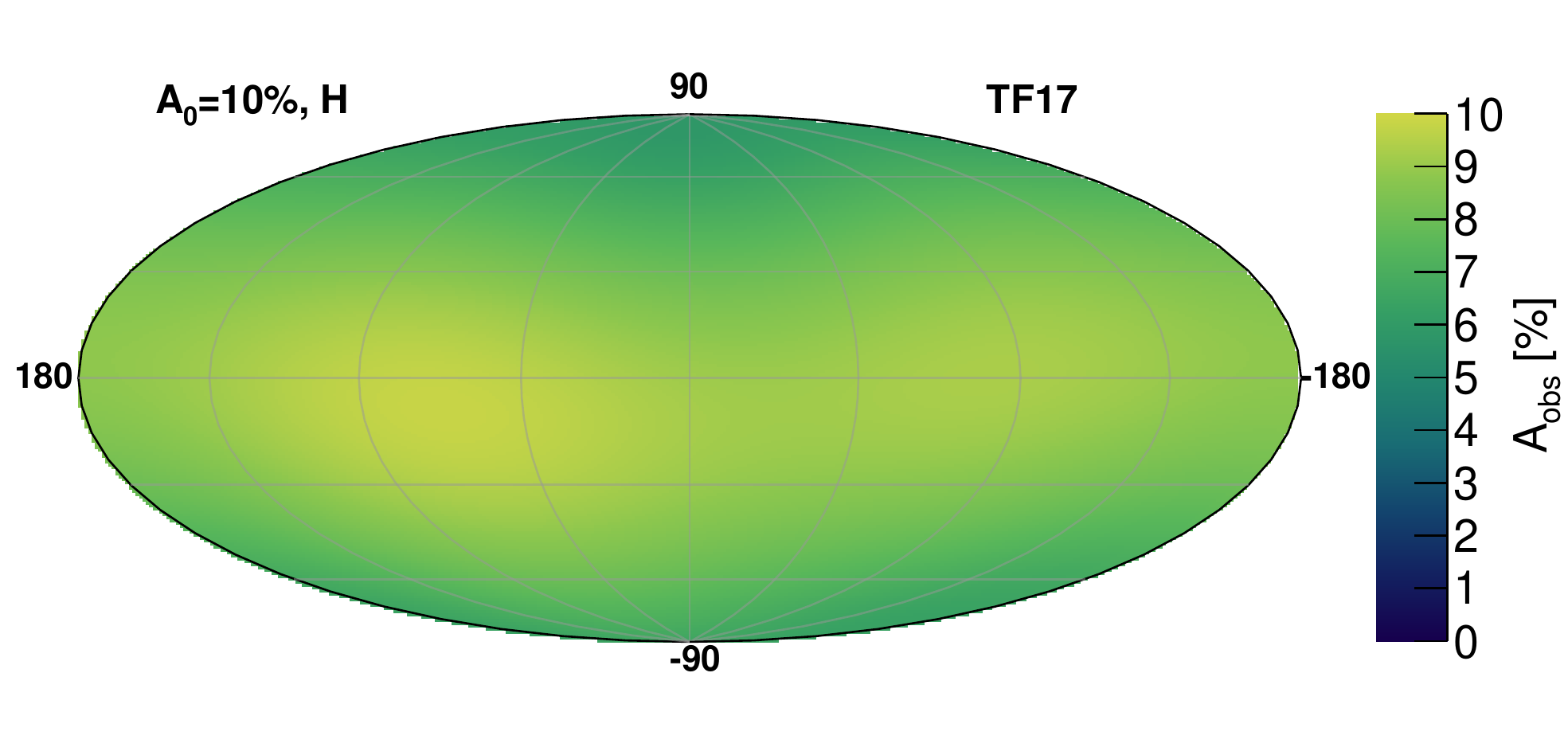}   
        
         \label{fig:amplitudeH_TF17}
     \end{subfigure}
     \begin{subfigure}{0.49\textwidth}
         \centering
          \includegraphics[width=1.0\textwidth]{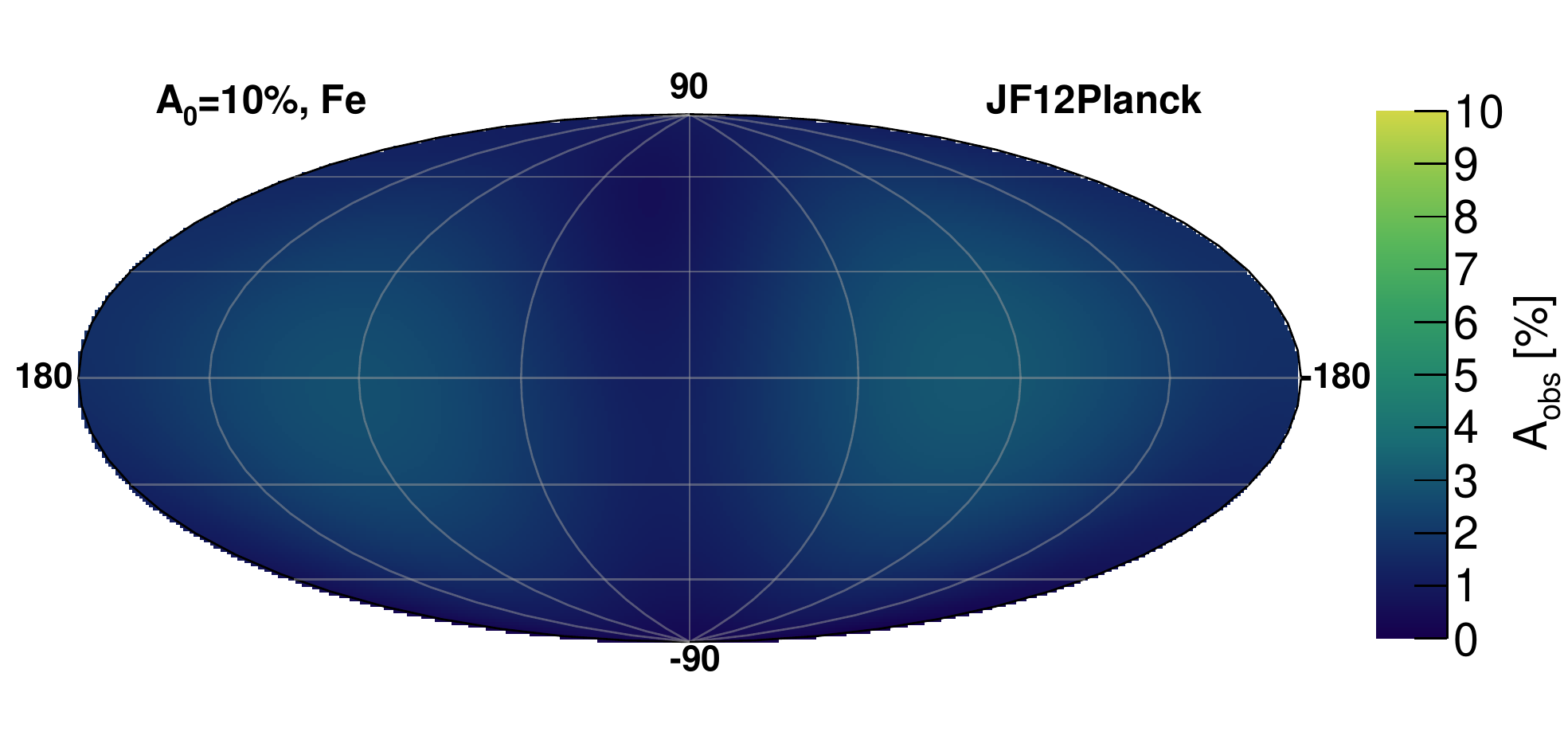}
         
         \label{fig:amplitudeFe_JF12}
     \end{subfigure}
     \hfill
     \begin{subfigure}{0.49\textwidth}
         \centering
        \includegraphics[width=1.0\textwidth]{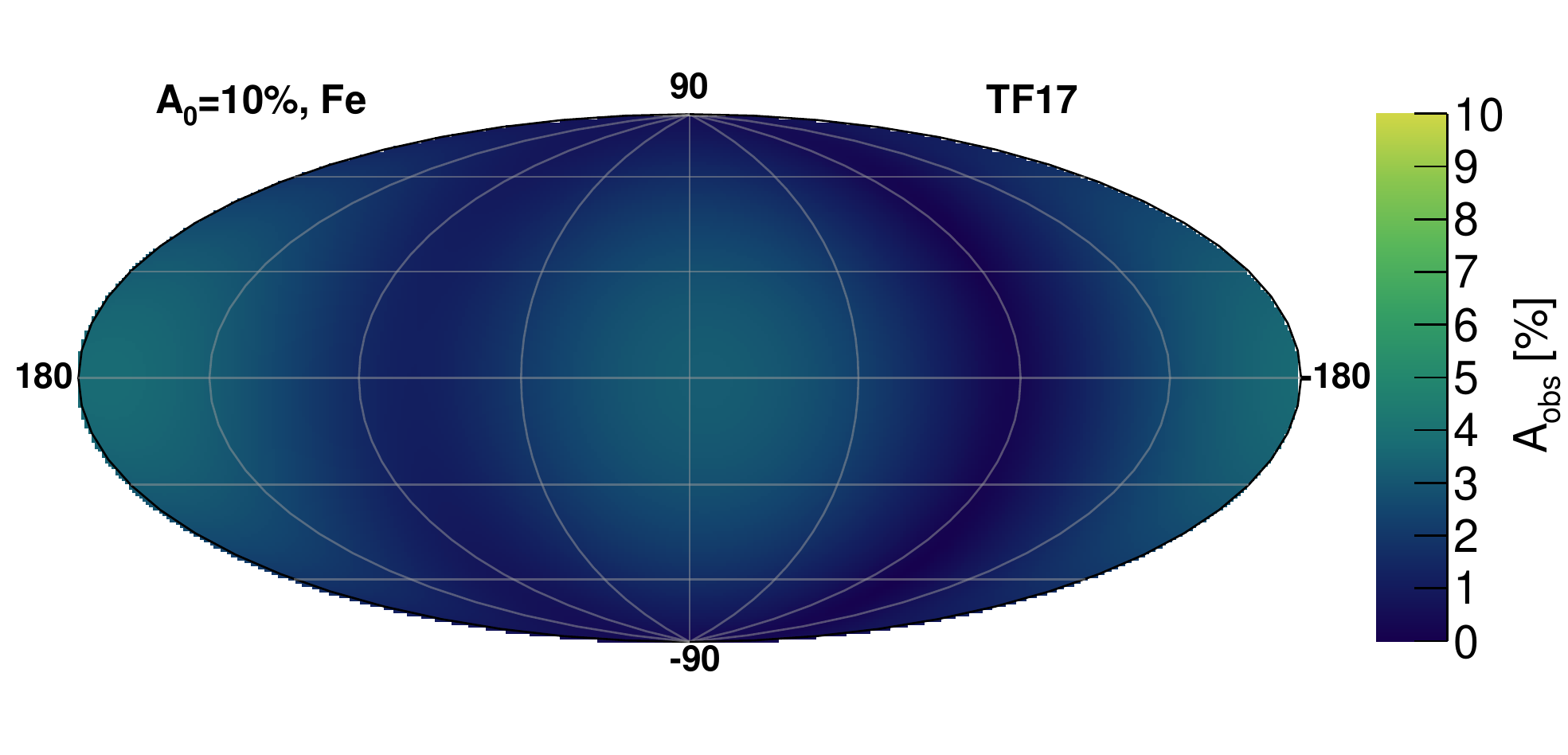}   
        
         \label{fig:amplitudeFe_TF17}
     \end{subfigure}
        \caption{Sky map in galactic coordinates of extragalactic dipole directions with initial amplitude of $A_0=10\%$ for protons (top) and iron nuclei (bottom) of energies above 8~EeV. The color scale corresponds to the amplitude of the dipole on the observer after propagation in the JF12Planck with $L_c=60$~pc (left) and the TF17 (Bd1C1) model of the GMF (right).}
        \label{fig:amplitudeH}
\end{figure}

\begin{figure}
     \centering
     \begin{subfigure}{0.49\textwidth}
         \centering
          \includegraphics[width=1.0\textwidth]{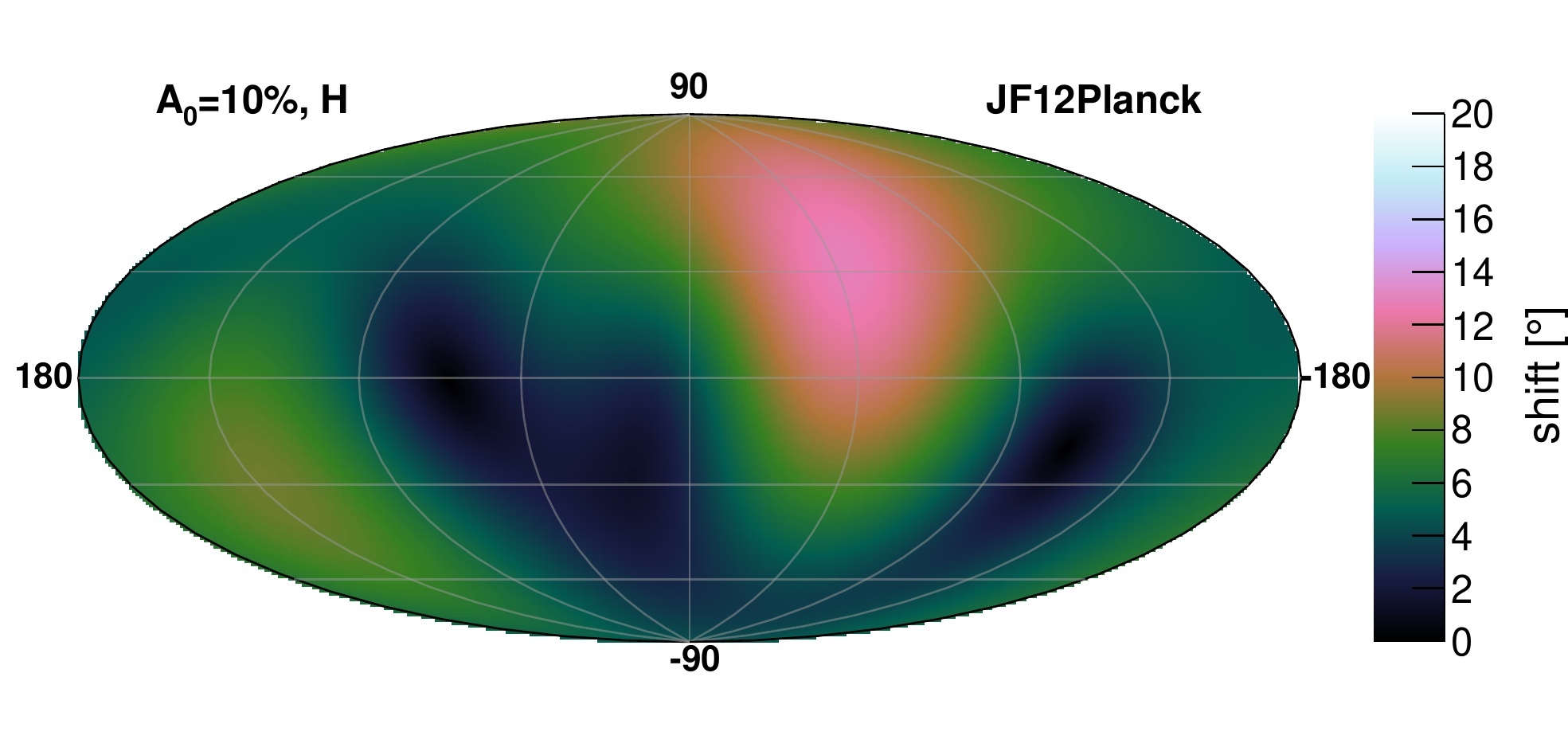}
         
         \label{fig:shiftH_JF12}
     \end{subfigure}
     \hfill
     \begin{subfigure}{0.49\textwidth}
         \centering
        \includegraphics[width=1.0\textwidth]{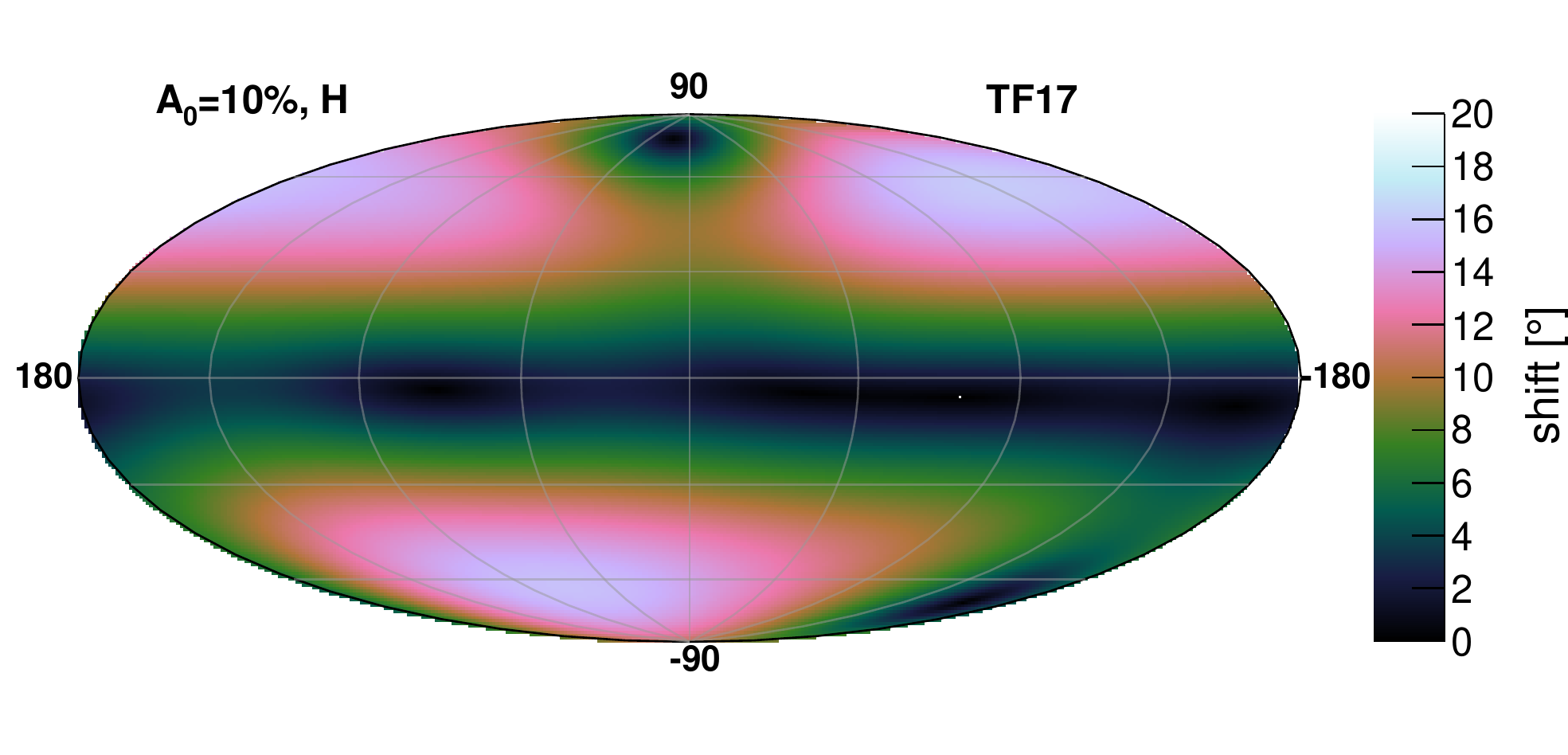}   
        
         \label{fig:shiftH_TF17}
     \end{subfigure}
  \begin{subfigure}{0.49\textwidth}
         \centering
          \includegraphics[width=1.0\textwidth]{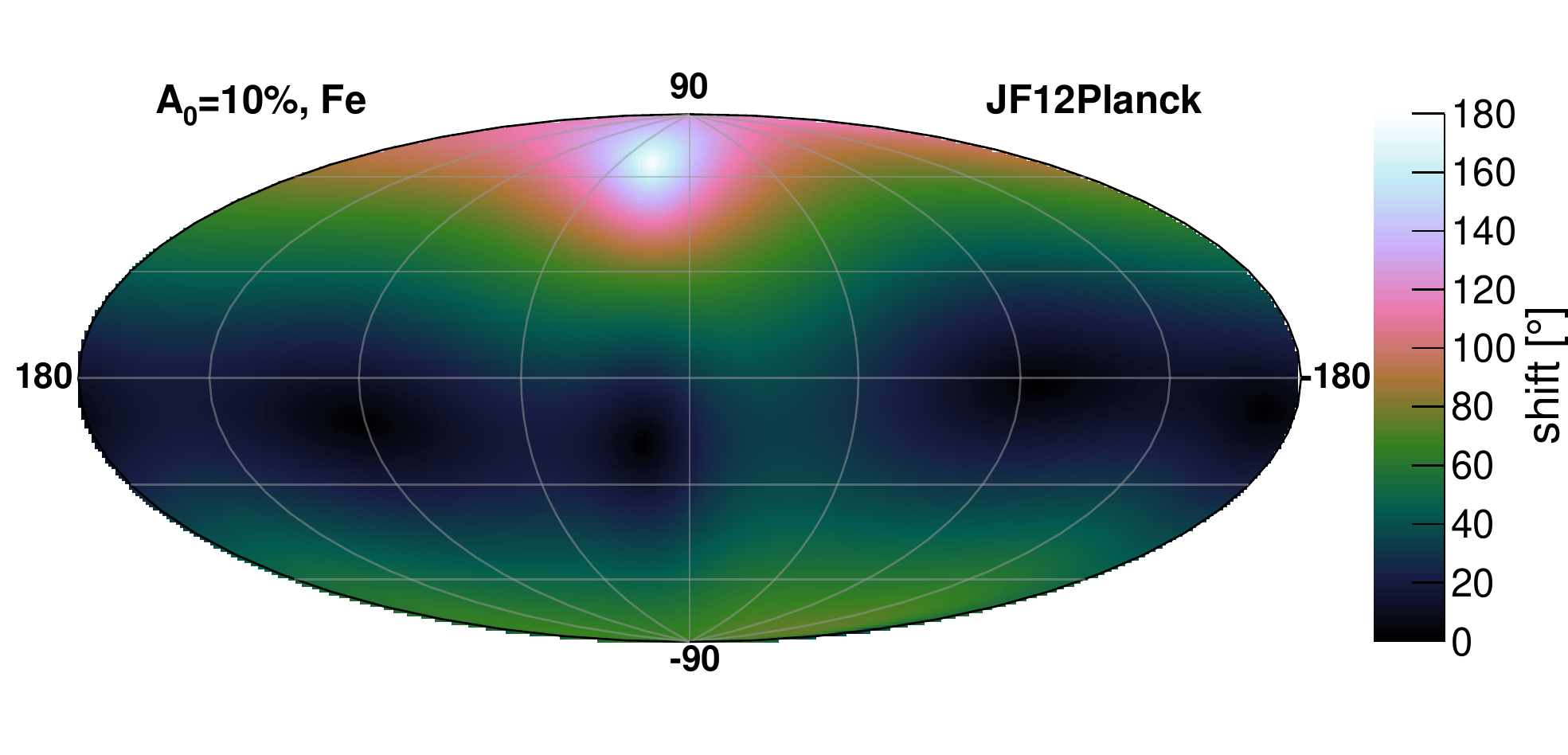}
         
         \label{fig:shiftFe_JF12}
     \end{subfigure}
     \hfill
     \begin{subfigure}{0.49\textwidth}
         \centering
        \includegraphics[width=1.0\textwidth]{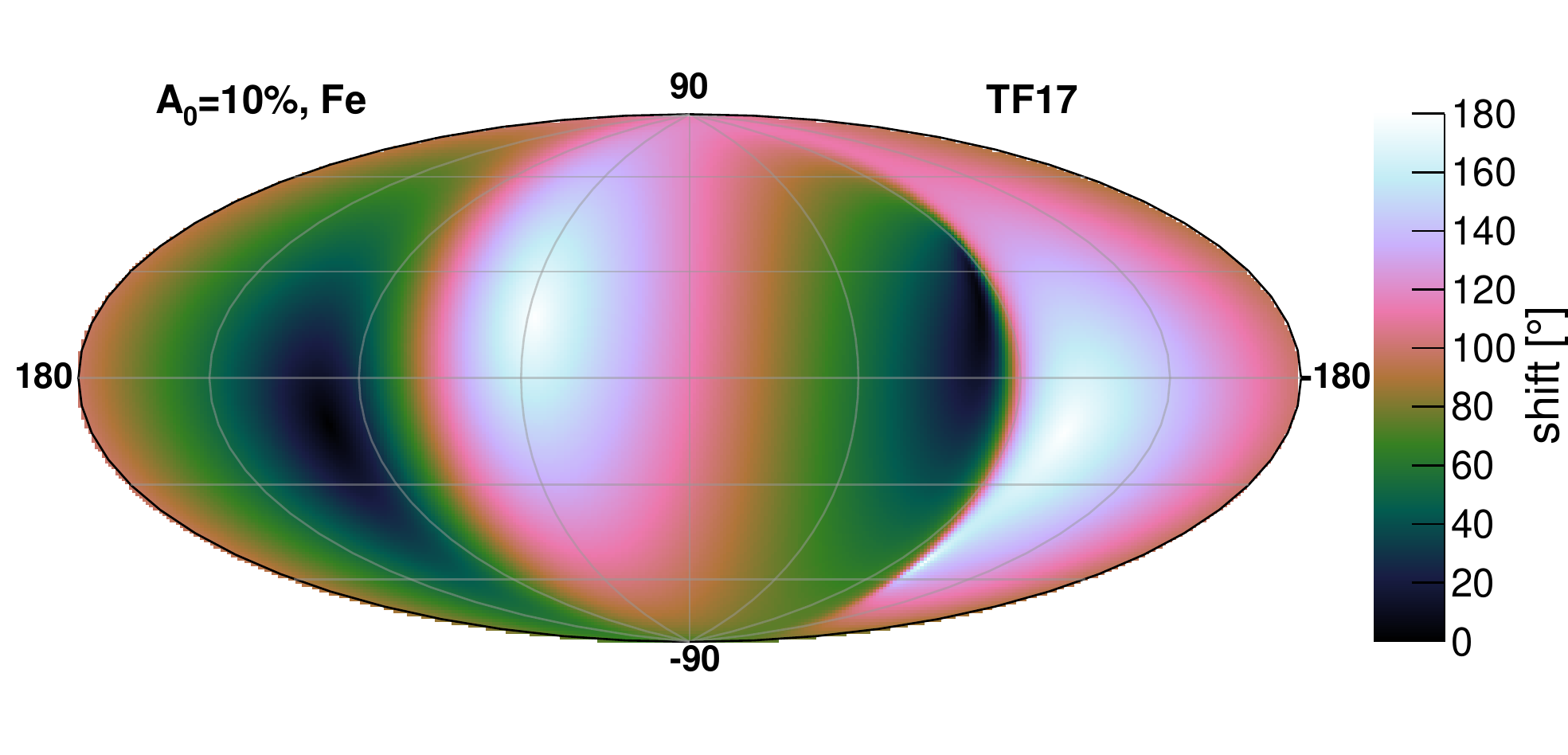}   
        
         \label{fig:shiftFe_TF17}
     \end{subfigure}
     
        \caption{The shift of the dipole direction for different extragalactic directions of the dipole for protons (top) and iron nuclei (bottom) of energies above 8~EeV using the JF12Planck model with $L_c=60$~pc (left) and the TF17 (Bd1C1) model of the GMF (right). The sky maps are in galactic coordinates.}
        \label{fig:dipoleShift}
\end{figure}

The effect of the GMF on the amplitude of the dipole is shown in Figure~\ref{fig:amplitudeH} for all combinations of longitude and latitude with a step of $1^{\circ}$ for proton and iron nuclei primaries with extragalactic amplitude 10\% for one realisation of the JF12Planck model of the GMF with coherence length $60$~pc and TF17 (Bd1C1) model of the GMF. The sky maps are in Galactic coordinates representing the sky above the observer and each bin corresponds to the respective extragalactic direction of the imposed dipole.  The color represents the amplitude of the dipole on the observer after the propagation in the two models of the GMF.

In case of the JF12Planck model, a minimal suppression of the amplitude on the observer is found in two extended lobes around longitudes of $\sim\pm90^{\circ}$ for both proton and iron nuclei, see Figure~\ref{fig:amplitudeH} (left). In the case of protons, the initial amplitude of the extragalactic dipole is suppressed maximally down to $\approx 8\%$ from the initial amplitude $A_0=10$\%, depending on the direction of the extragalactic dipole. 
However, the distribution of arrival directions of iron nuclei can get completely isotropised within the GMF for some directions of the extragalactic dipole. Therefore, the flux on the observer can be compatible with isotropic distribution. Similar suppression of the amplitudes were found for other realisations of the JF12Planck field.

Concerning the Bd1C1 option of the TF17 model, Figure~\ref{fig:amplitudeH} (right), different suppression behavior is seen for protons and for iron nuclei. In case of protons, minimal suppression of the amplitude is found for directions of the extragalactic dipole around the Galactic plane, while for iron nuclei we see minimal suppression of the amplitude in regions towards and opposite the Galactic center. The maximal suppression from the initial amplitude $A_0=10$\% is down to $\approx 6\%$ for protons. Similarly to the JF12Planck model of the GMF, complete isotropisation of the arrival directions can occure for some directions of the extragalactic dipole in case of iron nuclei for the TF17 model of the GMF. Similar behavior was found for the options Ad1C1 and Dd1C1 of the TF17 model. However, the suppression of the amplitude is lower in case of Dd1C1 option of the field than for the other two options of the TF17 field.

\subsection{Dipole Direction}

As well as the amplitude of the dipole is reduced after propagation in the GMF, the direction of the dipole can change too. Similarly to the previous case, the shift of the dipole direction depends on the model of the GMF used, rigidity of the particles and the extragalactic direction of the dipole. The shift can be in orders of few degrees, nevertheless, it can be as high as $\approx180^{\circ}$ in the case of heavy particles/low rigidities. 

To illustrate the possible shifts of the dipole direction, we show a sky map in galactic coordinates of the angular distance between the imposed extragalactic direction of the dipole and the reconstructed direction of the dipole on Earth after propagation in the GMF in Figure~\ref{fig:dipoleShift}. The figure shows the dipole shift for protons and iron nuclei propagated in one realisation of the JF12Planck model of the GMF with $L_c=60$~pc and in the Bd1C1 option of the TF17 model of the GMF. The skymap shows a sky above the observer and each bin represents an extragalactic direction of the dipole on a $1^{\circ}\times 1^{\circ}$ grid in longitude and latitude. The color represents the angular shift between the position of the extragalactic dipole and the direction of the dipole reconstructed on the observer. Note the different color scales that are used in order to better visualise the possible shifts for protons and iron nuclei. In the case of protons rather smaller angular shifts are predicted. The maximum angular distance between the dipole direction outside the Galaxy and at the observer for protons is found to be $\approx15^{\circ}$ using the JF12Planck model and $\approx 18^{\circ}$ for the Bd1C1 option of the TF17 model. However, taking iron nuclei propagated in the JF12Planck and Bd1C1 option of the TF17 models, in some directions of the extragalactic dipole the shift of the dipole can be much larger, going up to the aforementioned $180^{\circ}$.


\section{Compatibility with Observed Dipole Anisotropy}
\label{sec:results}

In order to find possible parameters of the extragalactic dipole that is observed on Earth by the Pierre Auger Observatory above $8$~EeV, we take into account both the amplitude and direction of the measured dipole. As the mass composition of cosmic rays is not known in large detail, we investigate all possible mass-composition mixes of the simulated four elements with a step of $5\%$ in relative fractions of the four types of particles, including single-component scenarios of p, He, N and Fe. The amplitude and direction of the dipole are reconstructed for individual mass-composition mixes, extragalactic directions of the dipole and extragalactic amplitudes using Equations~\eqref{eq:reco3d}. Parameters of the imposed amplitude and directions of the extragalactic dipole for a given mass-composition mix are selected as solutions if the reconstructed amplitude, as well as the direction of the dipole on the observer, are compatible with the measurement by the Pierre Auger Observatory within $1\sigma$ or $2\sigma$. All allowed directions of the extragalactic dipole were also checked to have a dipole behavior in the right ascension, see Appendix~\ref{C:fit}. In all the following figures, we show the allowed extragalactic directions for the JF12Planck as a combination of solutions obtained from all the different realisations of the GMF model. The same applies for the TF17 model of the GMF where the solutions are representing the combinations of the results obtained for the three options of the TF17 model, see Section~\ref{sec:sims}. 

\subsection{Single-Element Scenario}

Before we inspect the possible extragalactic directions of the dipole for all the mass-composition mixes, we start with results for the simple case of a single-element scenario. As the simulated energy spectrum is the same for all four types of particles, the influence of the GMF becomes more significant with higher atomic number as the rigidity is decreasing.

\begin{figure*}[htp]
     \centering
         \includegraphics[width=0.78\textwidth]{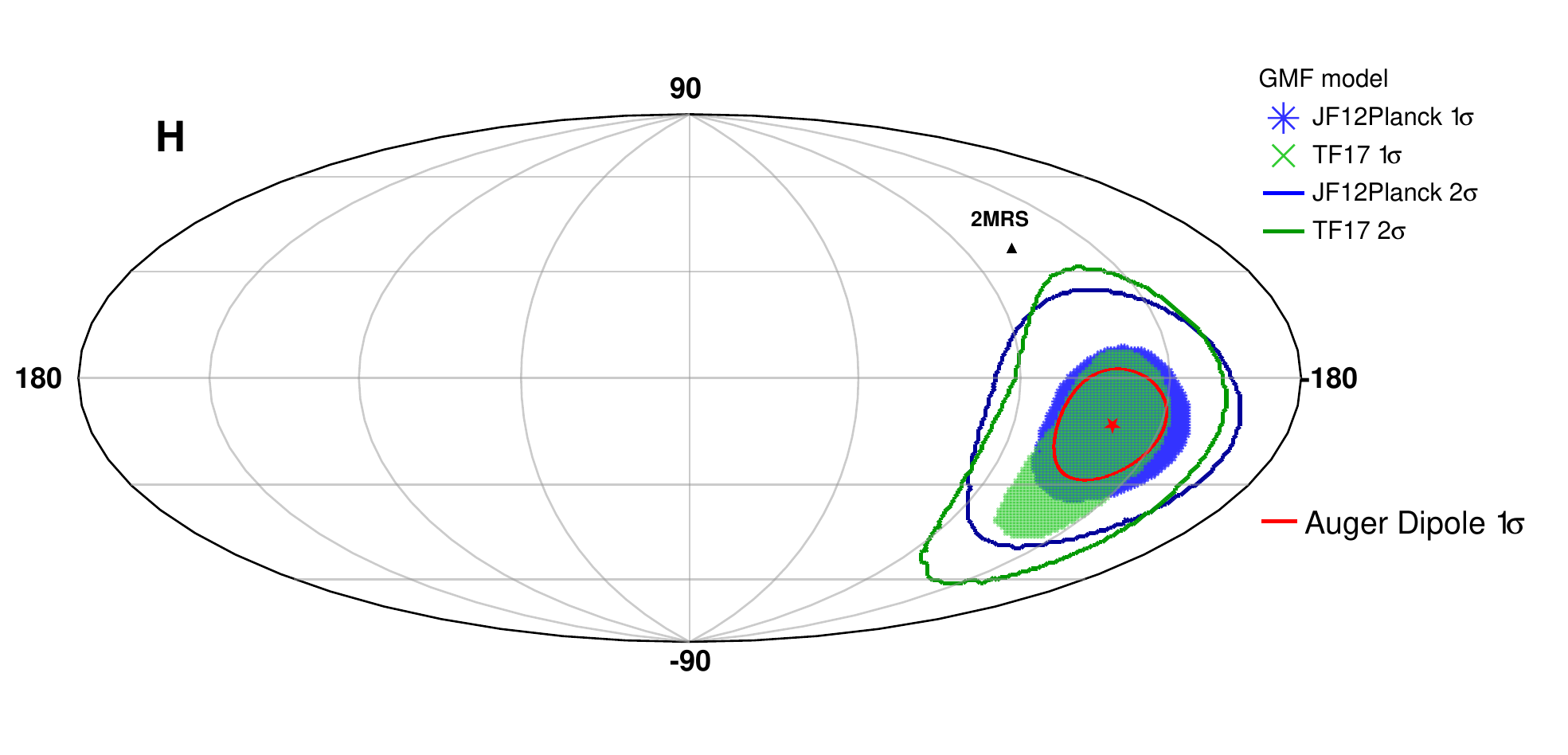}
        \caption{The directions of the extragalactic dipole compatible with the measured direction and amplitude by the Pierre Auger Observatory at $1\sigma$ and $2\sigma$ level found for the JF12Planck and the TF17 models of the GMF for the pure-proton scenario. The results from the Pierre Auger Observatory are indicated in red for 1$\sigma$ c.l. The plot is in Galactic coordinates. At the $1\sigma$ level the allowed extragalactic directions are found for initial amplitudes $A_0\leq10\%$ for both models of the GMF.}
        \label{fig:Hsingle}
\end{figure*}

\begin{figure*}[htp]
     \centering
         \includegraphics[width=0.78\textwidth]{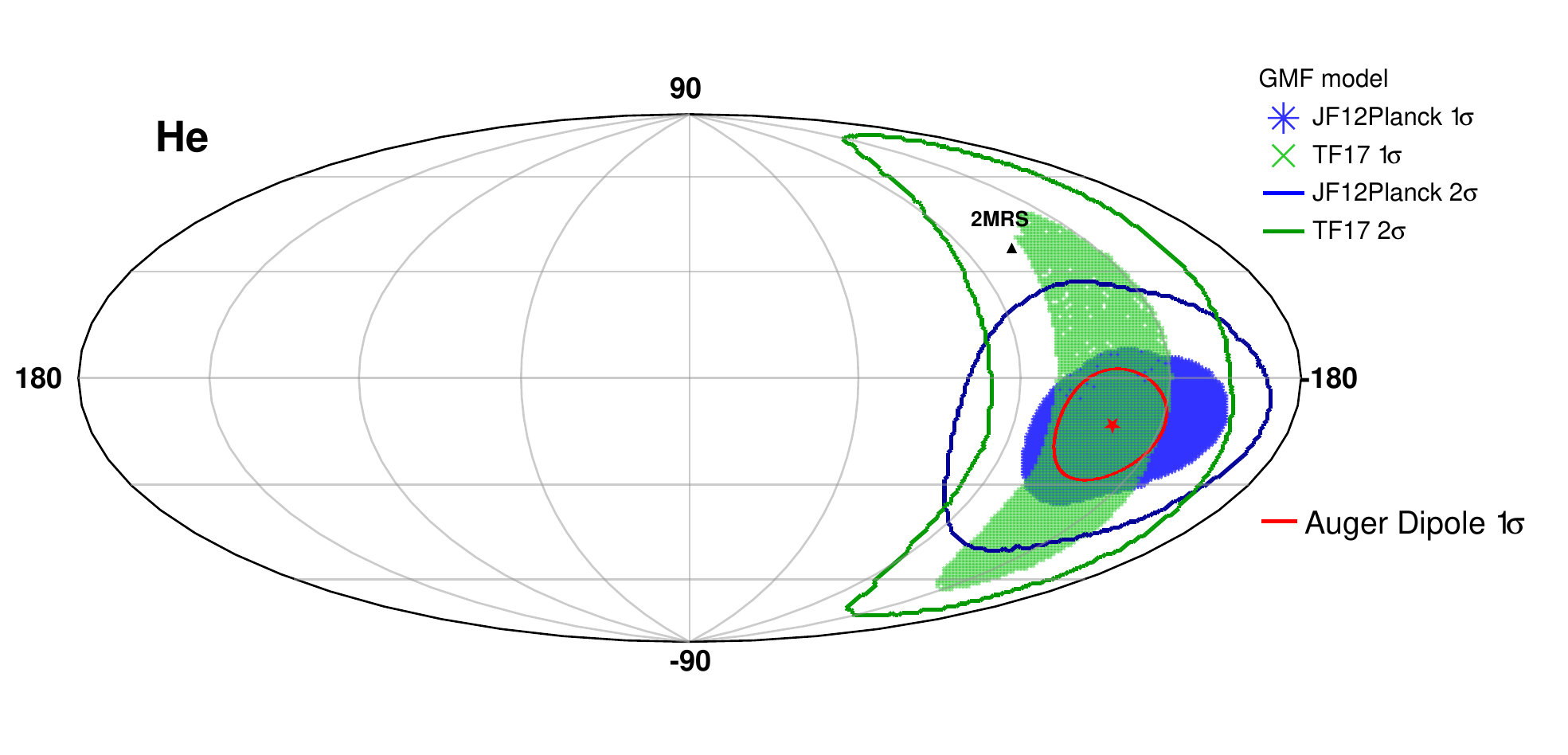}
        \caption{Same as in Figure~\ref{fig:Hsingle}, but for pure-helium scenario. At the $1\sigma$ level the allowed extragalactic directions are found for initial amplitudes $A_0\leq10\%$ and $A_0\leq16\%$ for JF12Planck and TF17, respectively.}
        \label{fig:Hesingle}
\end{figure*}

\begin{figure*}[hbt!]
     \centering
         \includegraphics[width=0.78\textwidth]{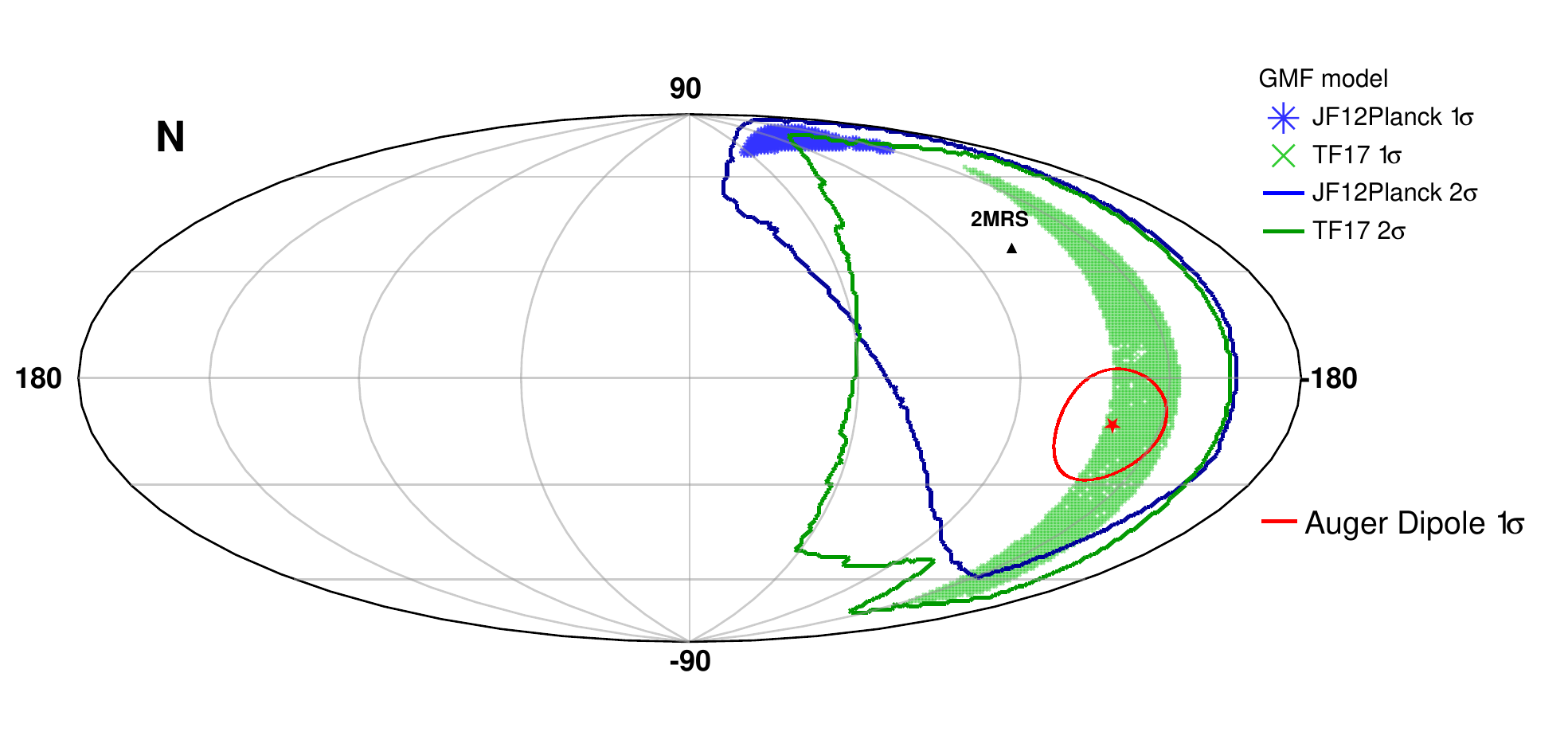}
        \caption{Same as in Figure~\ref{fig:Hsingle}, but for pure-nitrogen scenario. At the $1\sigma$ level the allowed extragalactic directions are found for initial amplitudes $A_0\geq14\%$ and $A_0\geq10\%$ for JF12Planck and TF17, respectively.}
        \label{fig:Nsingle}
\end{figure*}

\begin{figure*}[hbt!]
     \centering
         \includegraphics[width=0.78\textwidth]{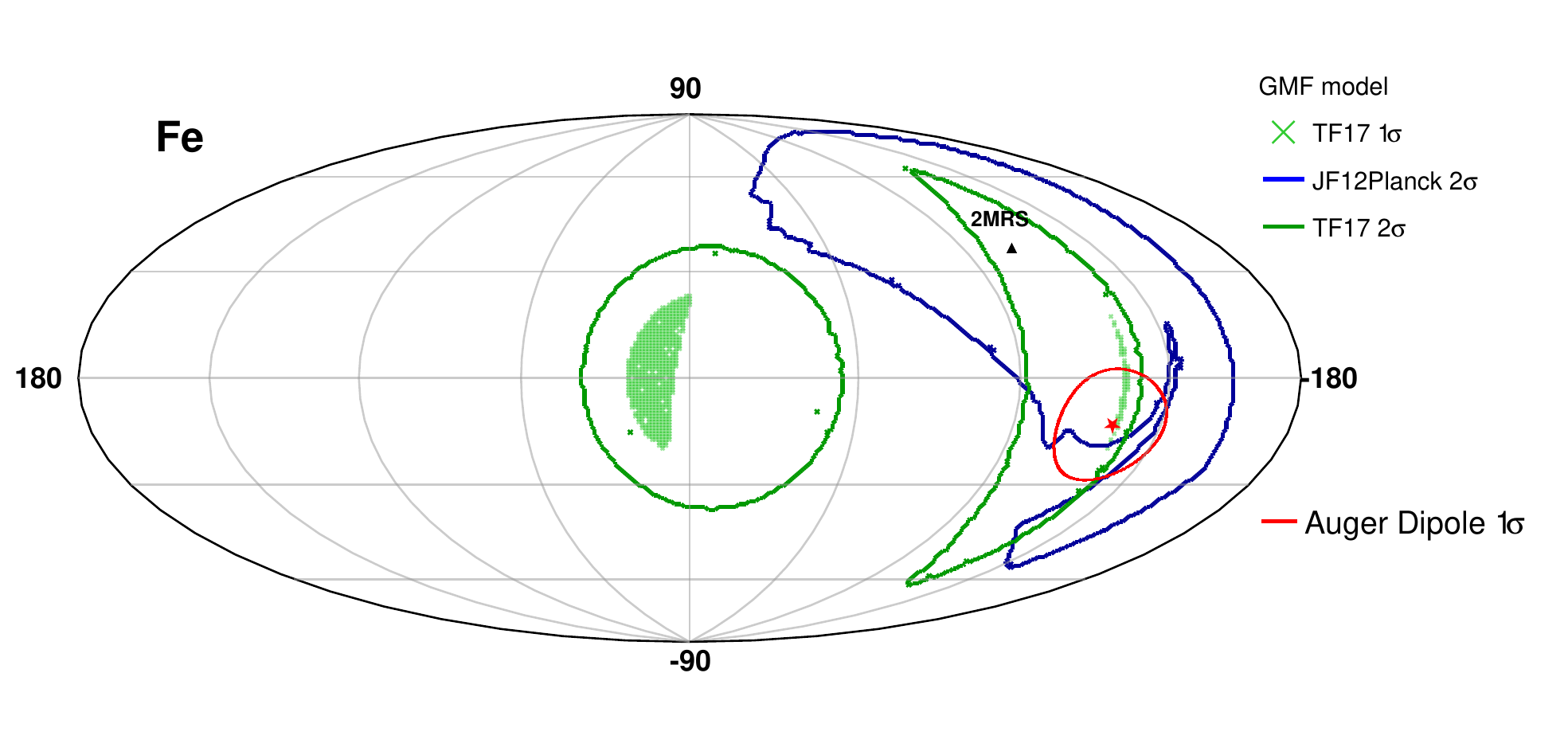}
        \caption{Same as in Figure~\ref{fig:Hsingle}, but for pure-iron scenario. No solutions are found for comparison at the $1\sigma$ level for the JF12Planck model. At the $1\sigma$ level the allowed extragalactic directions are found for initial amplitudes $A_0=20\%$ for the TF17 model of the GMF.}
        \label{fig:Fesingle}
\end{figure*}

At the $1\sigma$ level, we find that most of the possible directions of the extragalactic dipole can be found in the case of light elements; proton and helium, with an extragalactic amplitude lower or equal to $10\%$ for the JF12Planck model of the GMF and amplitude lower or equal to $16\%$ for the TF17 models of the GMF. The sky maps of the extragalactic directions of the dipole that are compatible with the measurements within 1$\sigma$ and $2\sigma$ are shown in Figure~\ref{fig:Hsingle} and Figure~\ref{fig:Hesingle} for pure protons and pure helium nuclei, respectively. All studied initial amplitudes of the extragalactic dipole are included. 
At the $1\sigma$ level and pure-proton scenario, the most distant solution is found within $\approx 25^{\circ}$ and $\approx 35^{\circ}$ from the measured direction of the dipole on Earth for the JF12Planck model and TF17 model of the GMF, respectively. In the case of pure-helium at the $1\sigma$ level, the maximum angular distance of an allowed extragalactic direction from the measured direction is $\approx35^{\circ}$ and $\approx60^{\circ}$ for the JF12Planck and the TF17 model, respectively. 

Possible extragalactic directions of the dipole for pure nitrogen scenario at the $1\sigma$ and $2\sigma$ level are depicted in Figure~\ref{fig:Nsingle} for the JF12Planck and the TF17 models of the GMF. In the case of the JF12Planck model of the GMF, extragalactic directions of the dipole at $1\sigma$ level are located far ($\approx 85^{\circ} - 105^{\circ}$) from the measured direction of the dipole on Earth with initial amplitudes $\geq 14\%$. For the TF17 model of the GMF at the $1\sigma$ level, the allowed extragalactic directions of the dipole are located in an extended band around longitudes from $210^{\circ}$ and $240^{\circ}$ and initial amplitudes \textcolor{blue}{$\geq 10\%$}. With the increasing initial amplitude the extragalactic directions of the dipole are allowed further away from the dipole direction measured by the Pierre Auger Observatory. Note that all the allowed directions of the extragalactic dipole for pure nitrogen scenario within the TF17 model are coming from the Dd1C1 option of the GMF model.  

The allowed extragalactic directions of the dipole for the pure iron nuclei scenario at the $1\sigma$ level and $2\sigma$ level are depicted in Figure~\ref{fig:Fesingle} for the JF12Planck and the TF17 models of the GMF. There are no allowed extragalactic directions of the dipole found for pure iron nuclei using the JF12Planck model at the $1\sigma$ level. However, two areas of allowed extragalactic directions are found in case of the TF17 model of the GMF at the $1\sigma$ level with initial amplitude $A_0=20\%$. 
The first group of allowed extragalactic directions is located in a narrow band around the mean longitude of $232^{\circ}$ and corresponds to solutions found using the Dd1C1 option of the TF17 model of the GMF. Second area of allowed extragalactic directions is found using the Bd1C1 option of the TF17 model of the GMF and is located close to the direction of the Galactic center. Note that this does not suggest a possible galactic origin of cosmic rays above 8~EeV as the simulated flux is of the extragalactic origin. At the $2\sigma$ level, we find possible directions of the extragalactic dipole for both models of the GMF.

\begin{figure*}[hbt!]
     \centering
         \includegraphics[width=\textwidth]{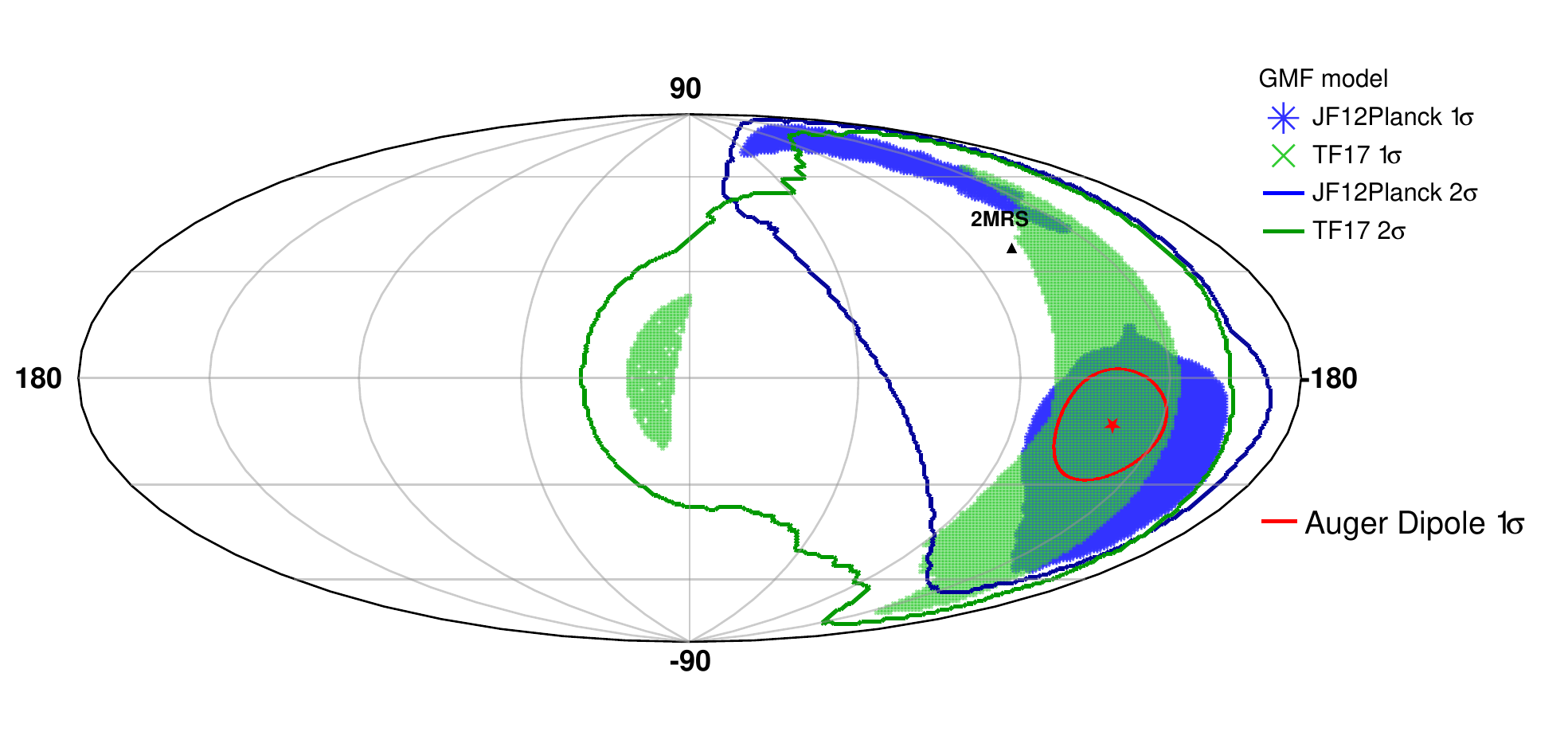}
        \caption{The directions of the extragalactic dipole in Galactic coordinates found for all various mass composition scenarios for the JF12Planck and TF17 models of GMF within $1\sigma$. Areas of possible directions of the extragalactic dipole compatible with the measurements within 2$\sigma$ are shown by blue and green contours for the JF12Planck and TF17 models, respectively. The 1$\sigma$ contour of the dipole measured by the Pierre Auger Observatory above 8 EeV is shown in red and the direction of the 2MRS dipole is displayed with a black triangle marker. Allowed directions of the extragalactic dipole are found for all investigated initial amplitudes $A_0$.}
        \label{fig:mixAll}
\end{figure*}

\subsection{Mixed Composition Scenario}
The experimental results regarding the mass composition of cosmic rays above $8$~EeV by the Pierre Auger Observatory suggest that the mass composition is mixed with the mean $\ln A$ evolving with the energy \cite{ICRC19_YushkovAuger, AugerXmax2014, AugerAnkle2016}. We assume that the abundance of individual elements remains the same in the whole energy range (due to the steeply falling energy spectrum, the dominant contribution to the large-scale anisotropy comes from the energies just above 8~EeV). Since we are not assuming any restrictions on the mass composition of cosmic rays, we are looking for allowed directions of the extragalactic dipole for all possible mass-composition mixes of the four elements. We mix the four particle species with a step in relative fractions of 5\%, going from the lightest, pure proton, to the heaviest, pure iron, composition. 

The possible directions of the extragalactic dipole for all mass-composition mixes, including the single-element scenarios, are shown in Figure~\ref{fig:mixAll} for both models of the GMF.  The JF12Planck model of the GMF allows most of the extragalactic directions at the 1$\sigma$ level in an extended area around the measured dipole direction, within $\approx45^{\circ}$. However, there is a second region of the allowed directions that goes as far as $\approx105^{\circ}$ from the measured direction of the dipole and these solutions are for nitrogen-dominated mass-composition mixes. The TF17 model suggests allowed directions of the extragalactic dipole around a narrow band of longitudes but a wide range of latitudes. Vast majority of the allowed extragalactic directions are within $\approx80^{\circ}$ from the measured direction of the dipole measured on Earth. The second region of allowed directions of the extragalactic dipole in a direction close to the Galactic center contains solutions for pure iron nuclei using the Bd1C1 option of the TF17 model of the GMF. The areas of allowed directions of the extragalactic dipole at the $2\sigma$ level extend much farther for both models of the GMF, specifically up to $\approx 115^{\circ}$ and $\approx 155^{\circ}$ from the measured direction of the dipole on Earth for the JF12Planck and TF17 model of the GMF, respectively.

\begin{figure*}[htp]
     \centering
         \includegraphics[width=1.0\textwidth]{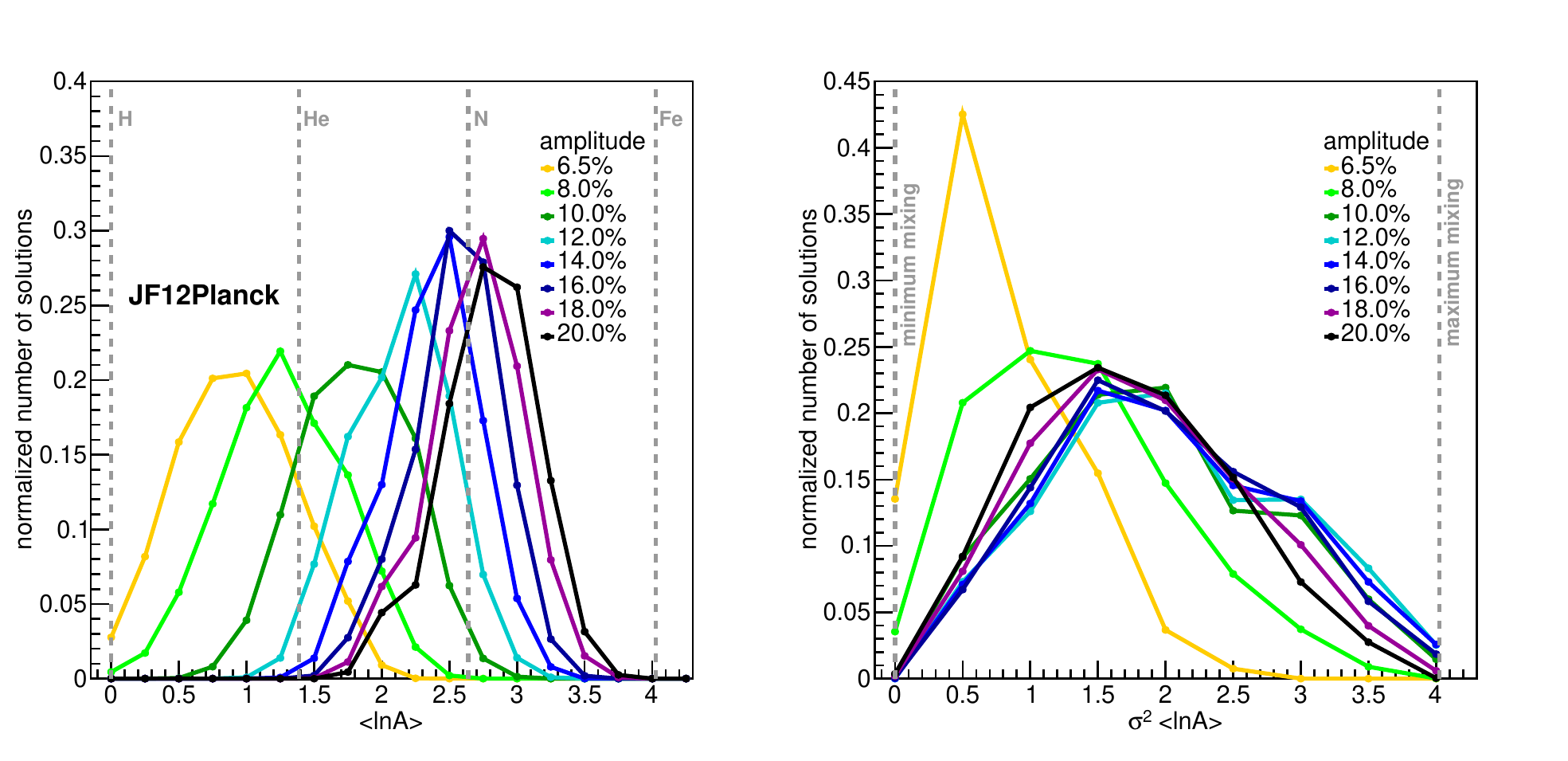}
        \caption{Number of allowed extragalactic directions of the dipole with respect to mean $\ln A$ (left) and with respect to the variance of the $\ln A$ (right) for different initial amplitudes for JF12Planck model of the GMF. Only directions with $1\sigma$ agreement with the measurement are shown. }
        \label{fig:mixAll_JF12}
\end{figure*}

\begin{figure*}[hbt!]
     \centering
         \includegraphics[width=1.0\textwidth]{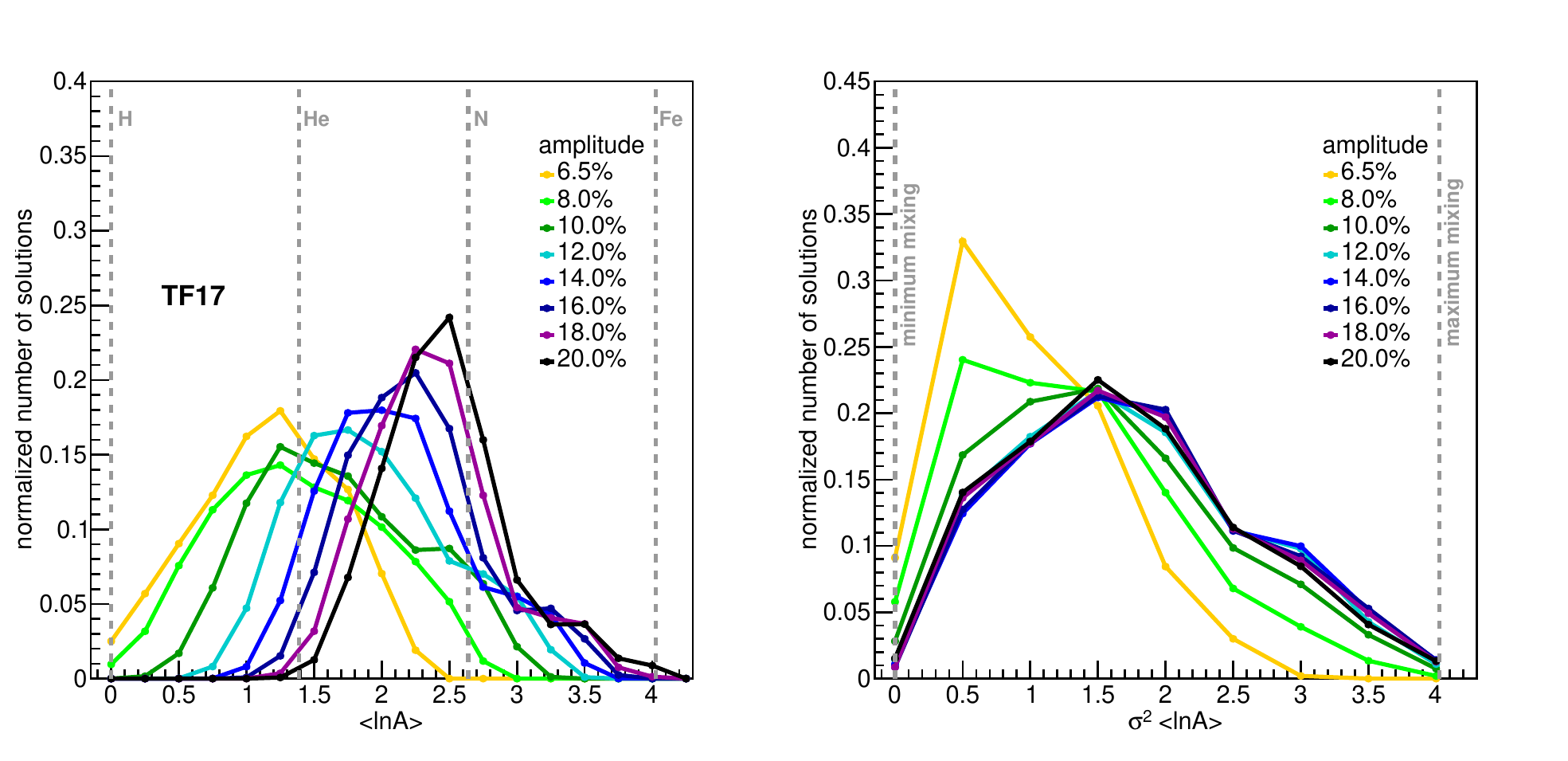}
        \caption{Same as in Figure~\ref{fig:mixAll_JF12}, but using the TF17 models of the GMF.}
        \label{fig:mixAll_TF17}
\end{figure*}

The normalized number of allowed directions of the extragalactic dipole in the $1^{\circ}$ by $1^{\circ}$ grid in longitude and latitude with respect to the mean (variance of) $\ln A$ for different initial amplitudes at the $1\sigma$ level is shown on the left (right) panels in Figure~\ref{fig:mixAll_JF12} and Figure~\ref{fig:mixAll_TF17} for the JF12Planck and TF17 models of the GMF, respectively. 
As expected, with a higher initial amplitude of the dipole outside the Galaxy a heavier composition is required to describe the data well. The heavy elements are needed in order to sufficiently suppress the amplitude of the dipole during propagation in the Galaxy for these high initial amplitudes.

\section{Discussion}
\label{sec:dis}

\textbf{General discussion:} We show the influence of the GMF on the dipole anisotropy using the JF12Planck and TF17 models of the GMF. These results are under the assumption of a dipole flux of extragalactic cosmic rays to the Galaxy. Different initial amplitudes $A_0$ of the dipole outside the Galaxy are allowed depending on the mean $\ln A$ of the cosmic-ray flux. Here, we investigate amplitudes from $6.5~\%$ up to $20~\%$ with discrete steps. Higher initial amplitudes of the dipole are not investigated as such high anisotropy in the cosmic-ray flux above 8~EeV would be difficult to explain by distribution of specific astrophysical sources.

The allowed directions of the extragalactic dipole of cosmic rays above 8 EeV that are presented in the previous section at the $1\sigma$ level mostly suggest directions within $\approx35^{\circ}$ (JF12Planck) and $\approx60^{\circ}$ (TF17) from the measured dipole direction on Earth for light elements and extending further from the measured dipole for heavier mass composition, going as far as $\approx 115^{\circ}$ and $\approx 155^{\circ}$ from the measured dipole on Earth at the $2\sigma$ level for the JF12Planck and TF17 models of the GMF, respectively. By including two models of the GMF and multiple realisations of each model we tried to include the uncertainties of the GMF models and make the results as general as possible. However, the results obtained with the TF17 model of the GMF should be interpreted with caution as this GMF model reaches very high amplitudes of the field strength, especially in the Ad1C1 and Bd1C1 options of the field. The obtained results were also investigated using ad-hoc adjustments of the strength of the TF17 model of the GMF (see Appendix~\ref{A:GMF}). 

Even though identification of sources creating the dipole anisotropy in arrival directions of cosmic rays above $8$~EeV is not the scope of this paper, we would like to mention that close to the allowed directions of the extragalactic dipole presented in the previous section is the direction of the 2MRS dipole \cite{2MRS}. Furthermore, at the $2\sigma$ level, the Centaurus~A region also lies within the region of allowed directions of the extragalactic dipole. The origin of large-scale anisotropies from the source distribution following a large-scale structure (LSS) was investigated in \cite{LSSdipole} and shows that a dipole similar to the one we observe on Earth might arise from such a source distribution. The authors find pure proton composition not to be compatible with the measured data in their scenario. However, in our results, we show that there exist allowed dipole directions for pure proton scenario. Note that these results are not in contradiction as in this work we allow the extragalactic dipole in all possible directions and not only towards the LSS. However, current measurements of UHECRs above 8~EeV are in contradiction with pure proton scenario \cite{ICRC19_YushkovAuger}.

The measurements of the Pierre Auger Observatory above $8$~EeV show the mean $\ln A$ of cosmic rays approximately heavier than helium nuclei \cite{ICRC19_YushkovAuger} with an indication of underestimation of the mean $\ln A$ due to uncertainties of models of hadronic interactions used for the mass-interpretation of measurements \cite{XmaxAdjust_Vicha}. Nevertheless, allowed directions of the extragalactic dipole with such $\ln A$ can be found for all the investigated initial amplitudes. Further constraints on the initial amplitude might be made using an energy dependent mass composition of cosmic rays, which is out of the scope of this current work. 

\bigskip

\textbf{Dipole at lower energies:} The measurements performed by the Pierre Auger Observatory show that the dipole anisotropy in the arrival directions of cosmic rays at lower energies, in the range from 4~EeV up to 8~EeV, has an amplitude of $2.5^{+1.0}_{-0.7}~\%$ and is pointing towards right ascension $(80\pm60)^{\circ}$ and declination $(-75^{+17}_{-8})^{\circ}$ \cite{AugerDipole2018}.  However, this dipole is not statistically significant ($<3\sigma$). 
Whether these two anisotropies have the same origin is not known as the distribution of sources of cosmic rays at energies ($4-8$)~EeV might be different than the distribution of sources of cosmic rays above $8$~EeV. Since the dipole at energies ($4-8$)~EeV has about three times lower amplitude, it is possible that additional sources, that are not capable of accelerating particles to the highest energies, contribute here as well, possibly even as an isotropic component. In order to investigate possible extragalactic directions of the dipole at lower energies, and the possibility of the same origin of the dipole anisotropies at these two energies, we repeat the same analysis as for the higher energies with simulations of cosmic rays with energies $(4-8)$~EeV. 

\begin{figure}[htp]
     \centering
         \includegraphics[width=0.9\textwidth]{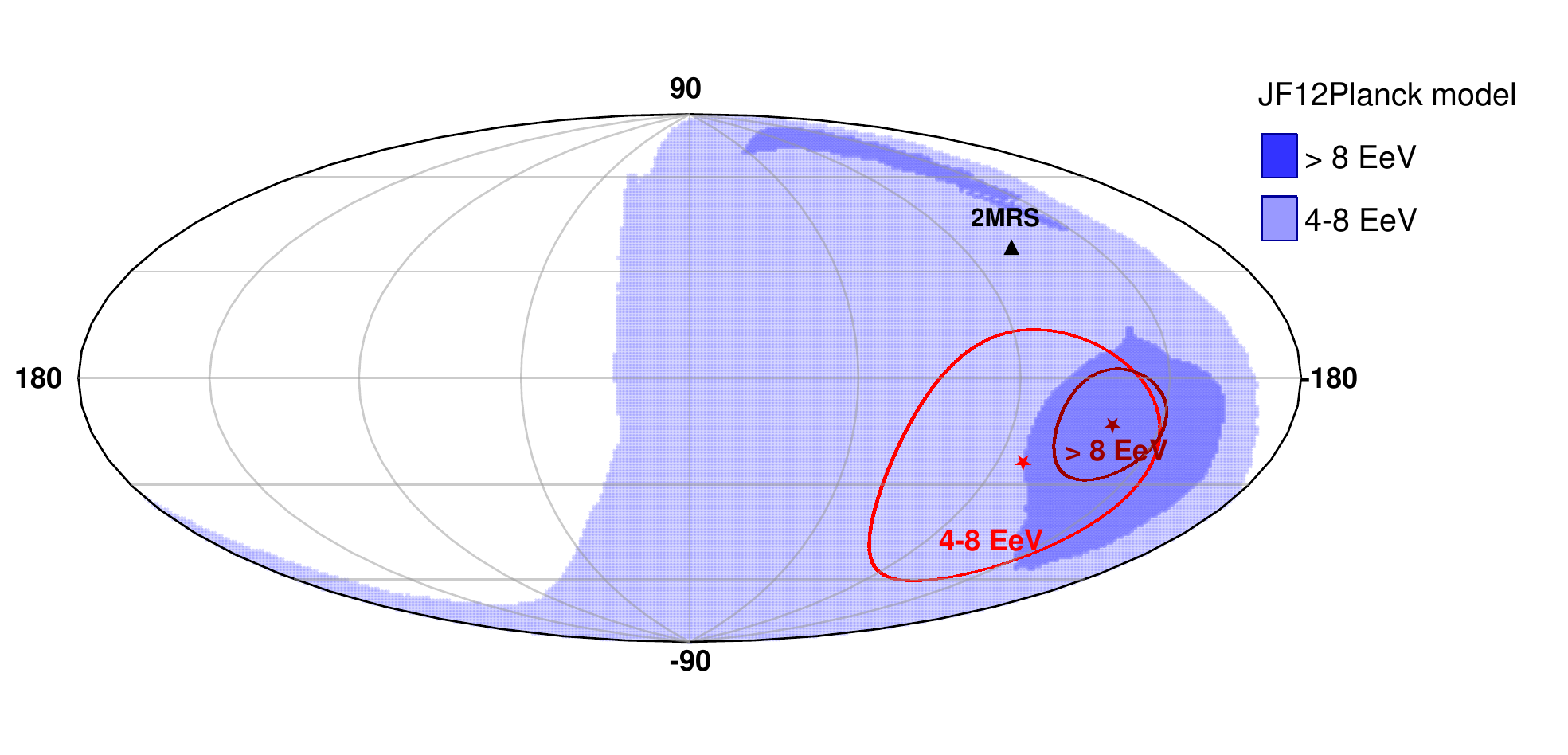}
        \caption{Areas of the allowed extragalactic dipole directions in galactic coordinates compatible with the measured direction of the dipole within $1\sigma$ by the Pierre Auger Observatory (red contour) with energies $4-8$ EeV (light blue) and above $8$~EeV (dark blue) using the JF12Planck model of the GMF. The amplitude of the lower energy dipole on Earth is considered to be less than 4\%.}
        \label{fig:lowE_JF12}
\end{figure}

\begin{figure}[htp]
     \centering
         \includegraphics[width=0.9\textwidth]{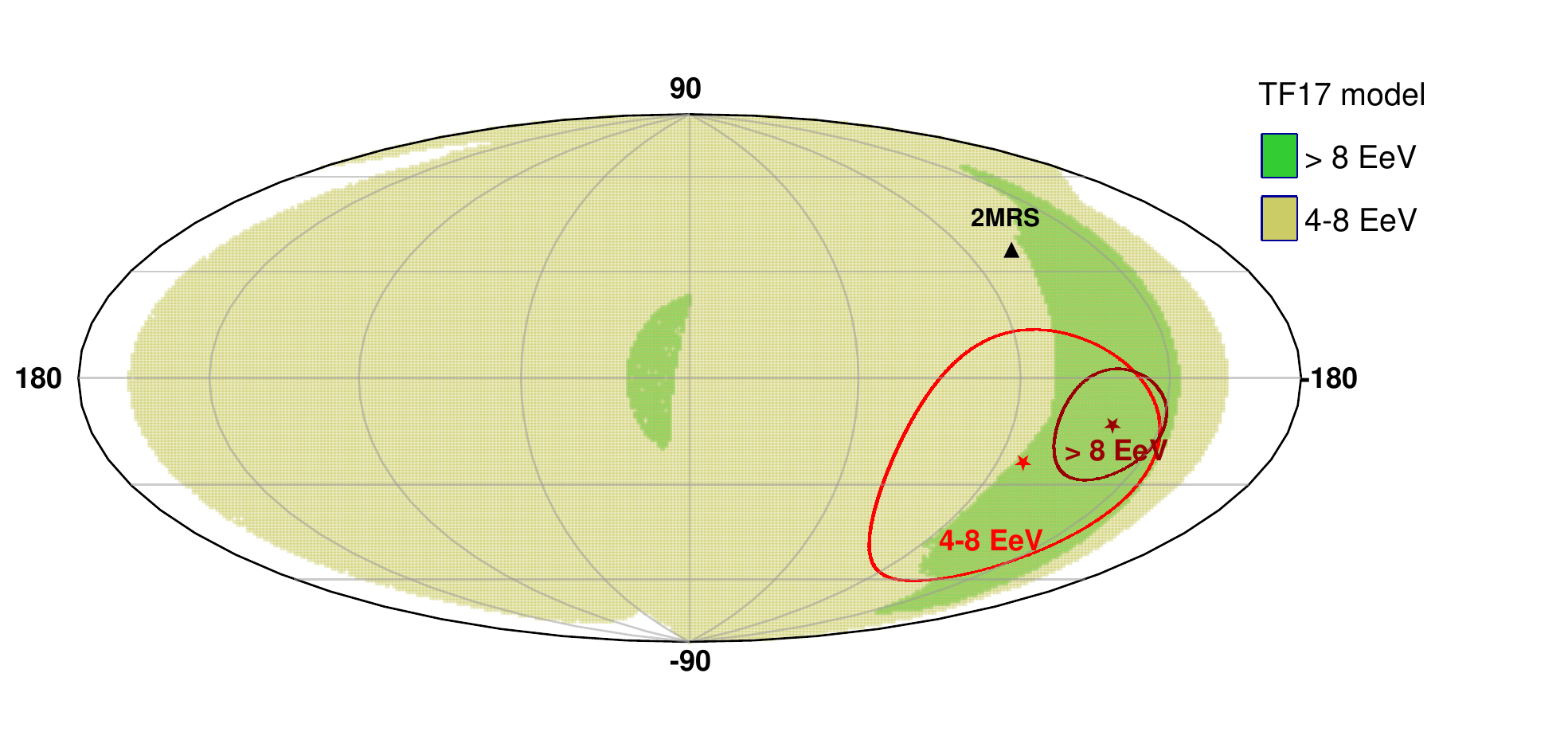}
        \caption{Same as in Figure~\ref{fig:lowE_JF12}, but using the TF17 model of the GMF.}
        \label{fig:lowE_TF17}
\end{figure}

We are searching for directions of the extragalactic dipole that would end up on the observer with amplitude and direction compatible at the $1\sigma$ level with the measurements by the Pierre Auger Observatory. The allowed directions of the extragalactic dipole for energies ($4-8$)~EeV and above $8$~EeV at the $1\sigma$ level are shown in Figure~\ref{fig:lowE_JF12} and Figure~\ref{fig:lowE_TF17} for the JF12Planck and TF17 models of the GMF, respectively.

 The regions of possible directions of the extragalactic dipole obtained from simulations at energies ($4-8$)~EeV are more extensive as the uncertainties of the measured dipole direction are much larger than for the dipole above $8$~EeV and the rigidities of particles are lower.  
 In case of the JF12Planck model of the GMF, the allowed extragalactic directions of the dipole at lower energies extend through more than half of the sky (Figure~\ref{fig:lowE_JF12}), mostly located at longitudes above $180^{\circ}$ through the whole range in latitude, while the TF17 model of the GMF allows possible extragalactic directions of the lower energy dipole almost across the whole sky (Figure~\ref{fig:lowE_TF17}). Nevertheless, the regions of allowed directions of the extragalactic dipole at lower energies and above $8$~EeV are overlapping for both models of the GMF. However, given the extensive areas of allowed directions of the extragalactic dipole, we can not confirm or reject the same origin of these two anisotropies. The dipole at lower energies might be further suppressed by additional sources, which is not taken into account here.

\section{Conclusions}
\label{sec:conc}

We investigate the influence of the Galactic magnetic field (GMF) on the dipole anisotropy of ultra-high-energy cosmic rays (UHECRs) that is measured by the Pierre Auger Observatory \cite{AugerDipole2017}. In our study, we assume an extragalactic origin of the UHECRs above 8~EeV with a dipole distribution in arrival directions when entering the Milky Way. Using simulations of the UHECR propagation in two models of the GMF, JF12Planck with three different coherence lengths of the turbulent filed and three options of the TF17 model of the GMF, we estimate the effects on the amplitude and direction of the dipole anisotropy in the arrival directions of UHECRs.

The comparison of the simulations with the measurement of the dipole is performed at the $1\sigma$ and $2\sigma$ level. At the $1\sigma$ level, rather low initial amplitudes of the dipole outside the Galaxy are needed ($\leq10\%$) in the case of light composition (mostly protons and helium nuclei) and the JF12Planck model of the GMF. In case of the TF17 model of the GMF, initial amplitudes up to 14\% are allowed for light composition. Higher initial amplitudes ($10\%-20\%$) result in compatible dipoles in arrival directions for heavier composition mixes for both models of the GMF. Most of the allowed extragalactic directions of the dipole are within $\approx45^{\circ}$ (JF12Planck), resp. $\approx80^{\circ}$(TF17) from the dipole direction observed on Earth. At the $2\sigma$ level, the sky regions of allowed directions of the extragalactic dipole are much more extended going to allowed directions of the extragalactic dipole as far as $\approx 115^{\circ}$ in case of the JF12Planck and $\approx155^{\circ}$ for the TF17 model of the GMF from the measured direction on Earth.

\appendix

\section{Galactic magnetic field models}
\label{A:GMF}
\begin{figure}[htp]

 \centering
     \begin{subfigure}{0.49\textwidth}
         \centering
          \includegraphics[width=1.0\textwidth]{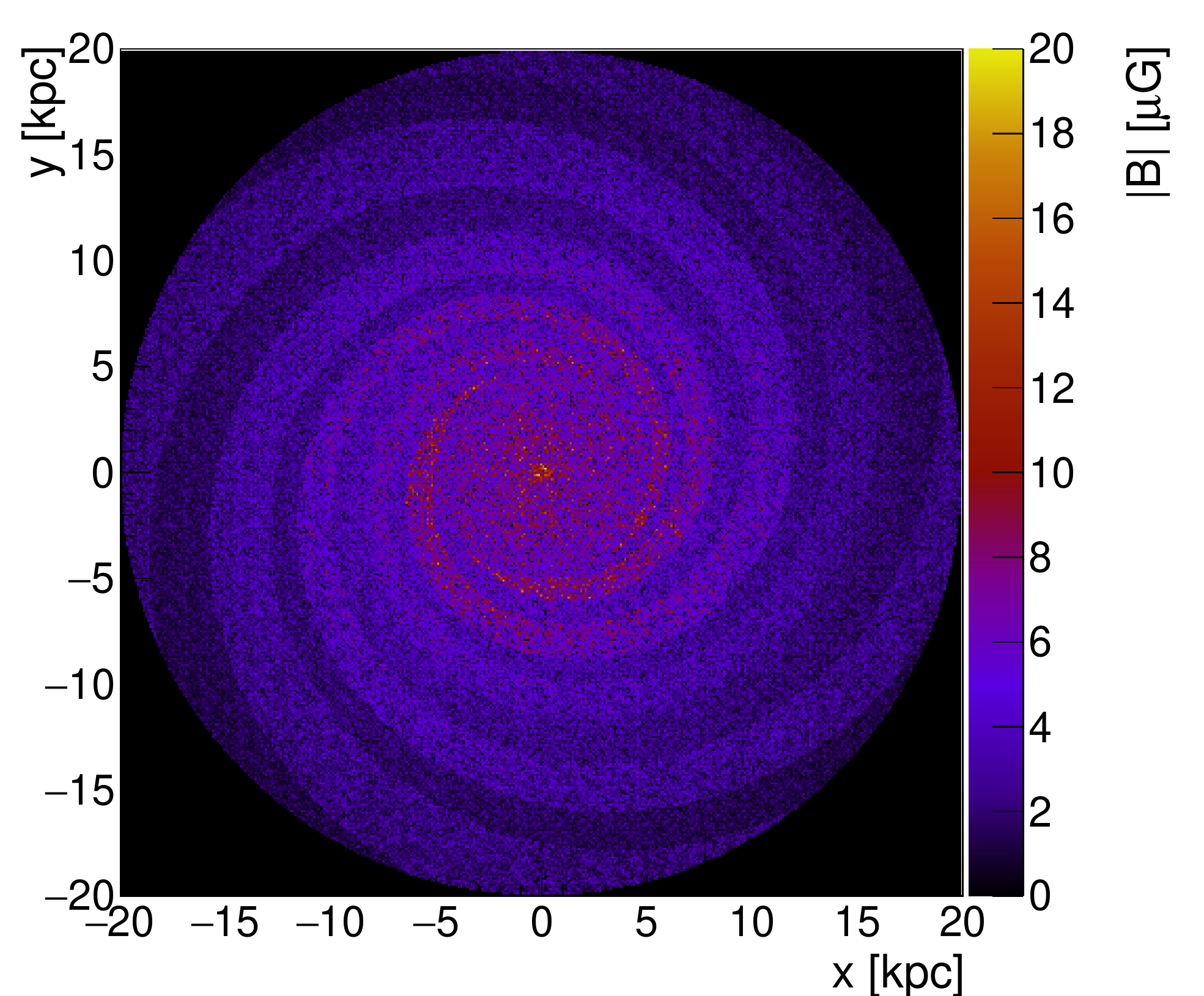}
         
         \label{fig:JF12XY}
     \end{subfigure}
     \hfill
     \begin{subfigure}{0.49\textwidth}
        \centering
        \includegraphics[width=1.0\textwidth]{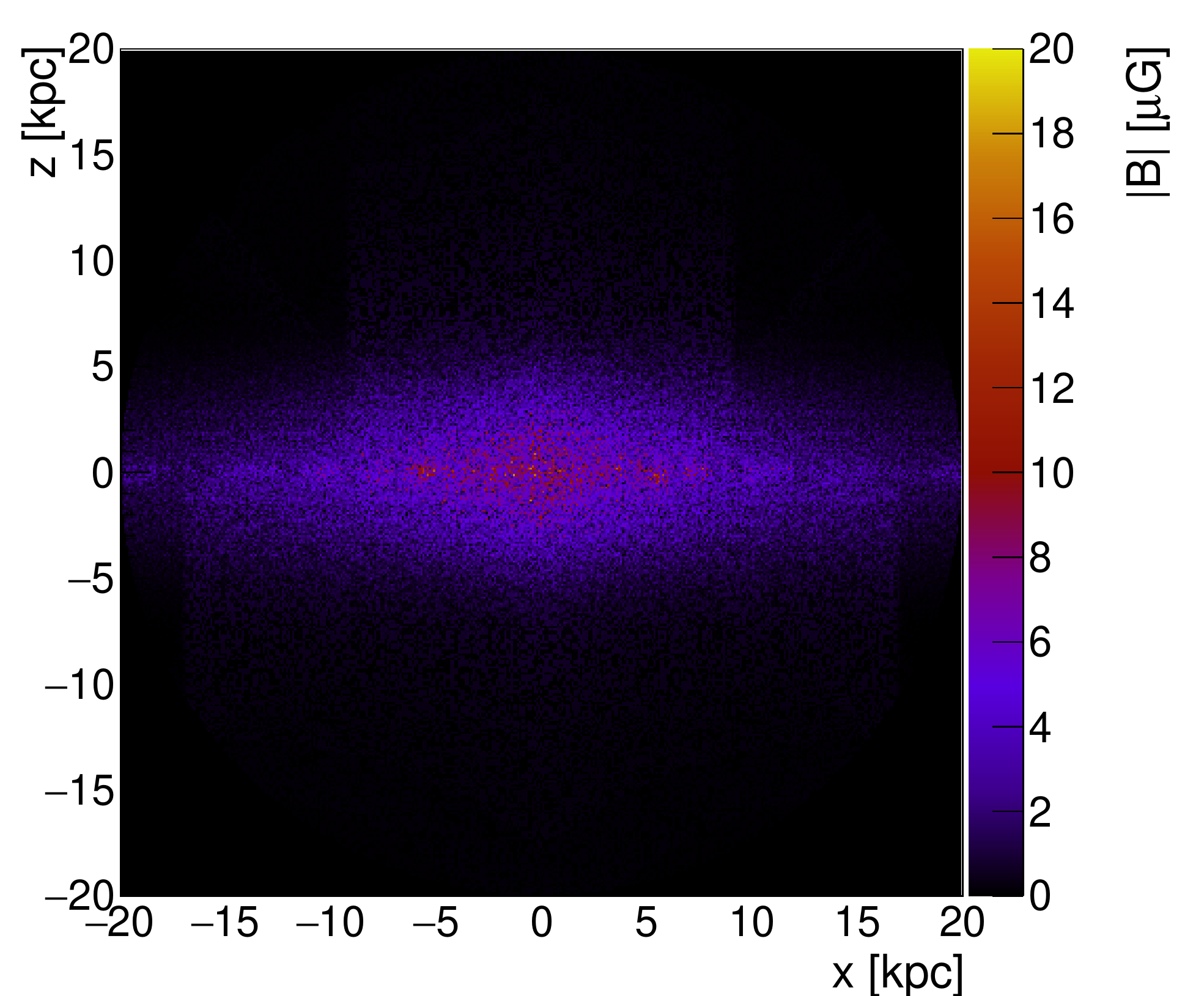}   
        
         \label{fig:Jf12XZ}
     \end{subfigure}
     \caption{Vizualisation of the strength of the JF12Planck model of GMF field in $xy$ plane (left) and $xz$ plane (right) with the Galactic center in the coordinates $(x,y,z)=(0,0,0)$~kpc.}
        \label{fig:GMFJF12Planck}
\end{figure}

The JF12Planck field composes of coherent, large-scale random (striated) and small-scale random (turbulent) fields \cite{JF12, JF12b}. The turbulent field of the JF12Planck model of the GMF is modeled within CRPropa~3 as Kolmogorov-type turbulent magnetic field \cite{CRPropa3}. We simulate particle propagation for three coherence lengths of the turbulent field, namely $L_c=30$~pc, $L_c=60$~pc and $L_c=100$~pc. Lowering the coherence length has a similar effect as reducing the field strength \cite{LSSdipole}. For each coherence length, we perform multiple realisations of the random turbulent field using different random seeds. A visualisation of one realisation of the field for $L_c=60$~pc in $xy$ plane at $z=0$~pc and in $xz$ plane at $y=0$~pc is shown in Figure~\ref{fig:GMFJF12Planck}.

\begin{figure}
  \begin{subfigure}{0.49\textwidth}
         \centering
          \includegraphics[width=1.0\textwidth]{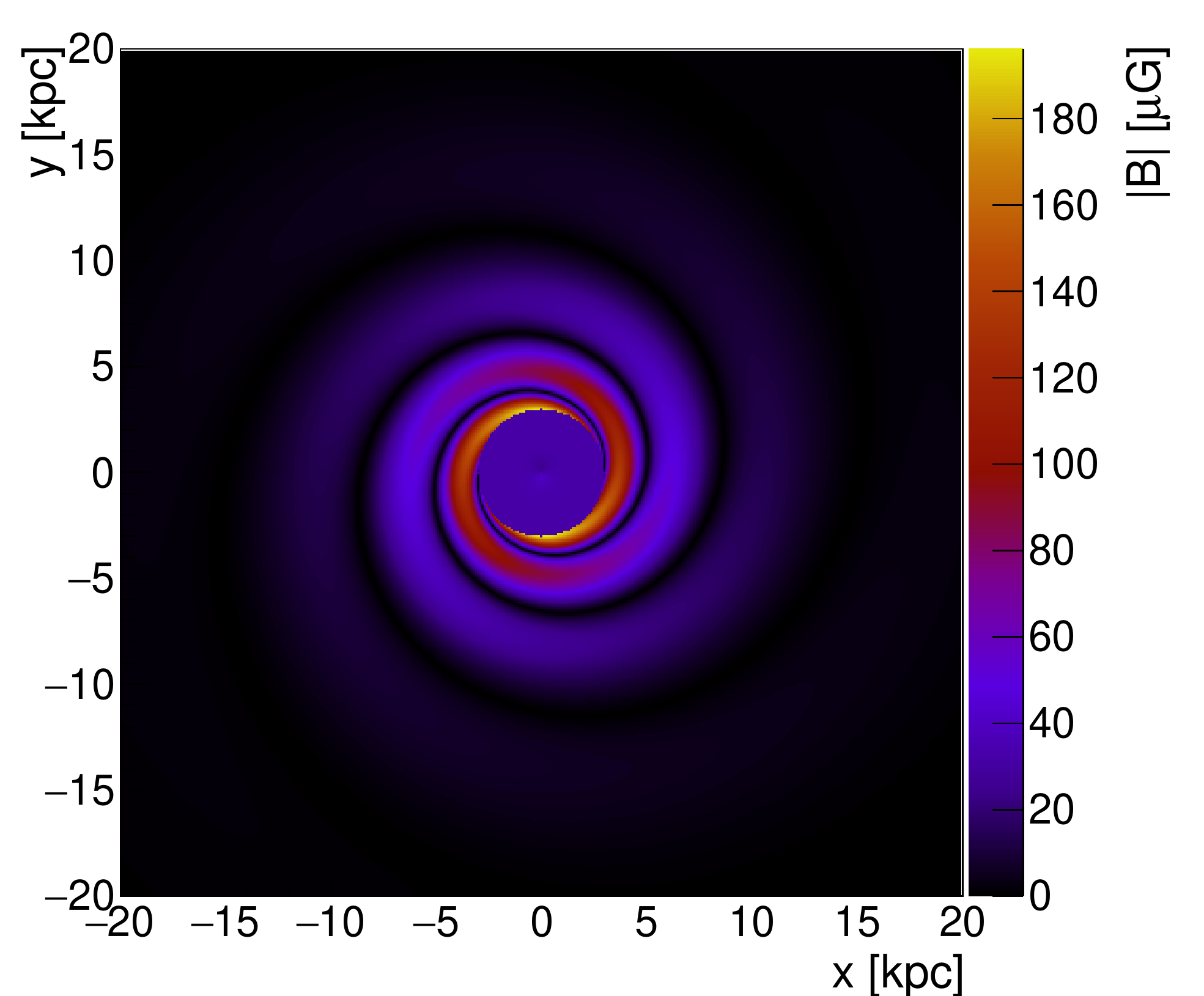}
         
         \label{fig:Ad1C1XY}
     \end{subfigure}
     \hfill
     \begin{subfigure}{0.49\textwidth}
         \centering
         \includegraphics[width=1.0\textwidth]{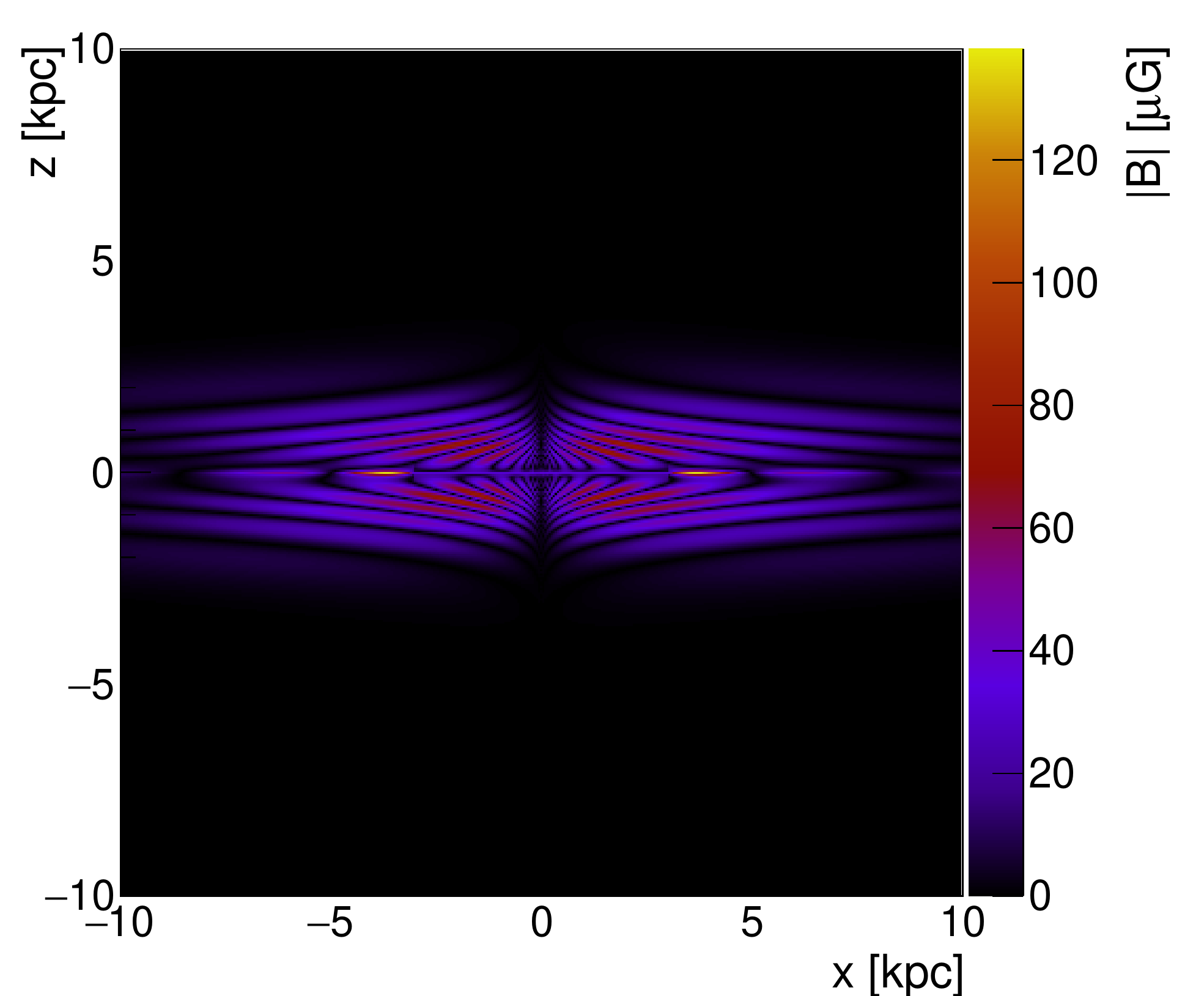}  
         
         \label{fig:Ad1C1XZ}    
     \end{subfigure}
     \caption{Same as Figure~\ref{fig:GMFJF12Planck} but for the Ad1C1 variant of the TF17 model of the GMF.}
    \label{fig:GMFAd1}
\end{figure}

\begin{figure}

 \begin{subfigure}{0.49\textwidth}
         \centering
          \includegraphics[width=1.0\textwidth]{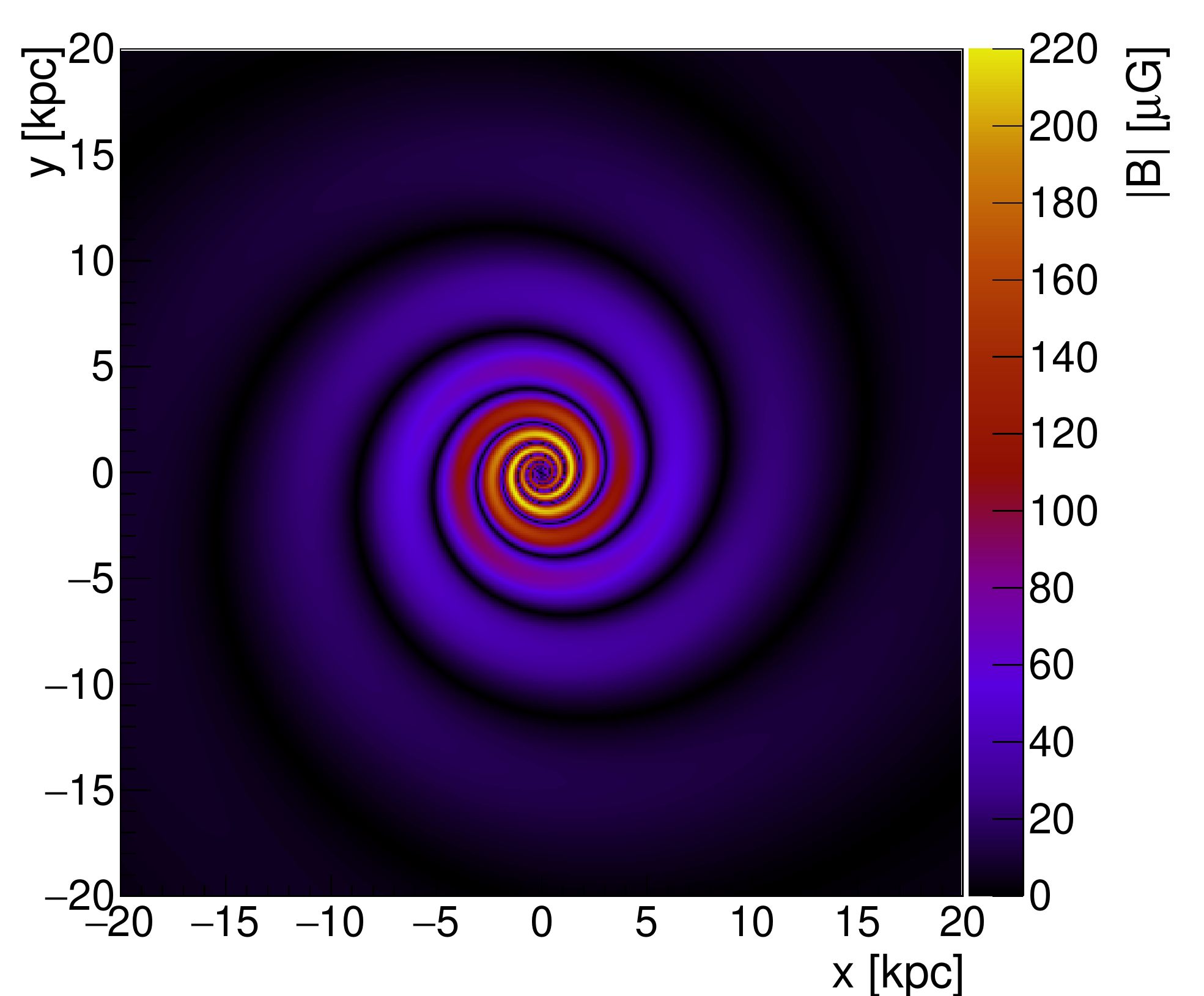}
         
         \label{fig:Bd1C1XY}
     \end{subfigure}
     \hfill
     \begin{subfigure}{0.49\textwidth}
         \centering
         \includegraphics[width=1.0\textwidth]{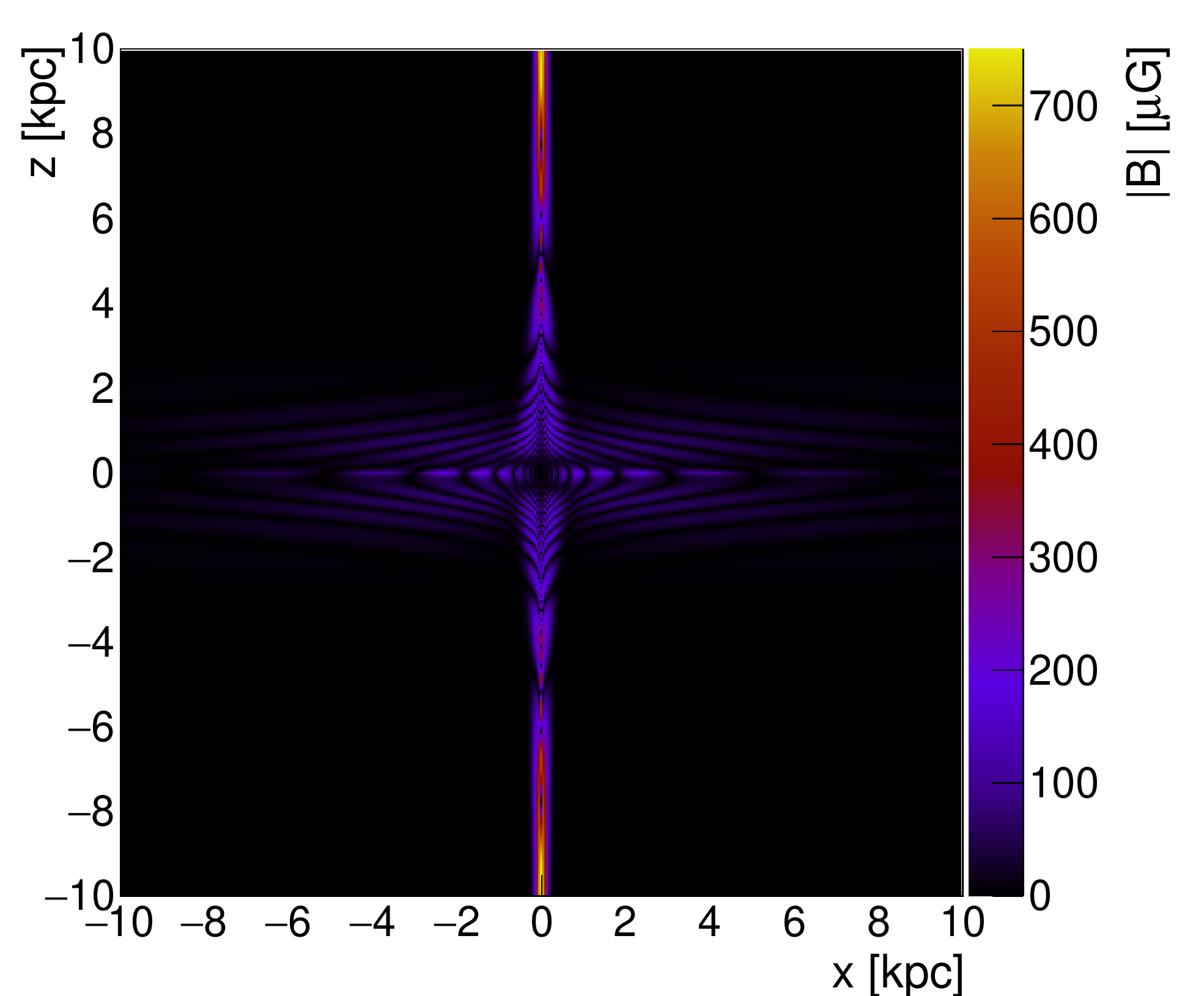}   
         
         \label{fig:Bd1C1XZ}    
     \end{subfigure}
     \caption{Same as Figure~\ref{fig:GMFJF12Planck} but for the Bd1C1 variant of the TF17 model of the GMF.}
    \label{fig:GMFBd1}
\end{figure}

\begin{figure}

      \begin{subfigure}{0.49\textwidth}
         \centering
          \includegraphics[width=1.0\textwidth]{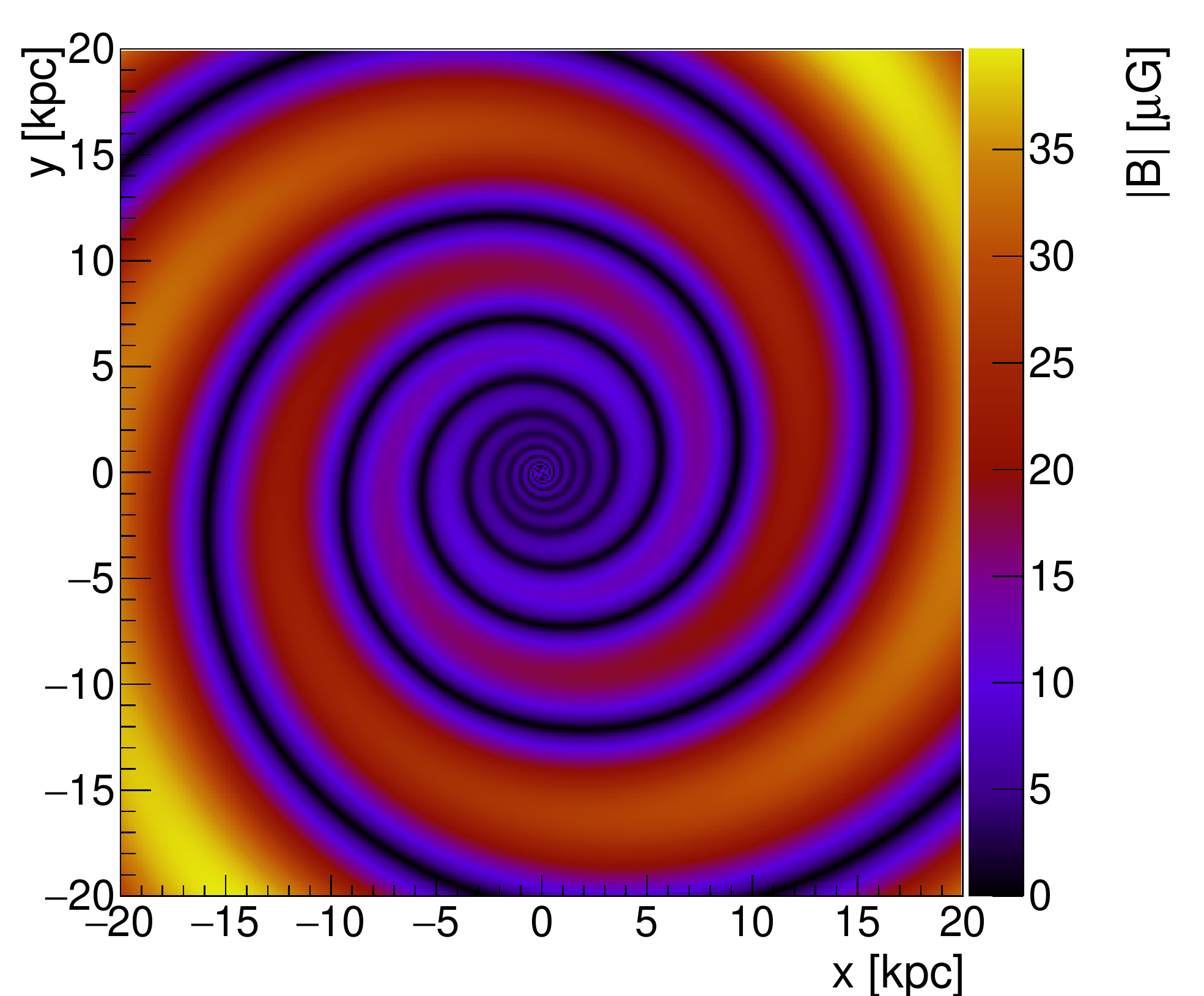}
         
         \label{fig:Dd1C1XY}
     \end{subfigure}
     \hfill
     \begin{subfigure}{0.49\textwidth}
         \centering
         \includegraphics[width=1.0\textwidth]{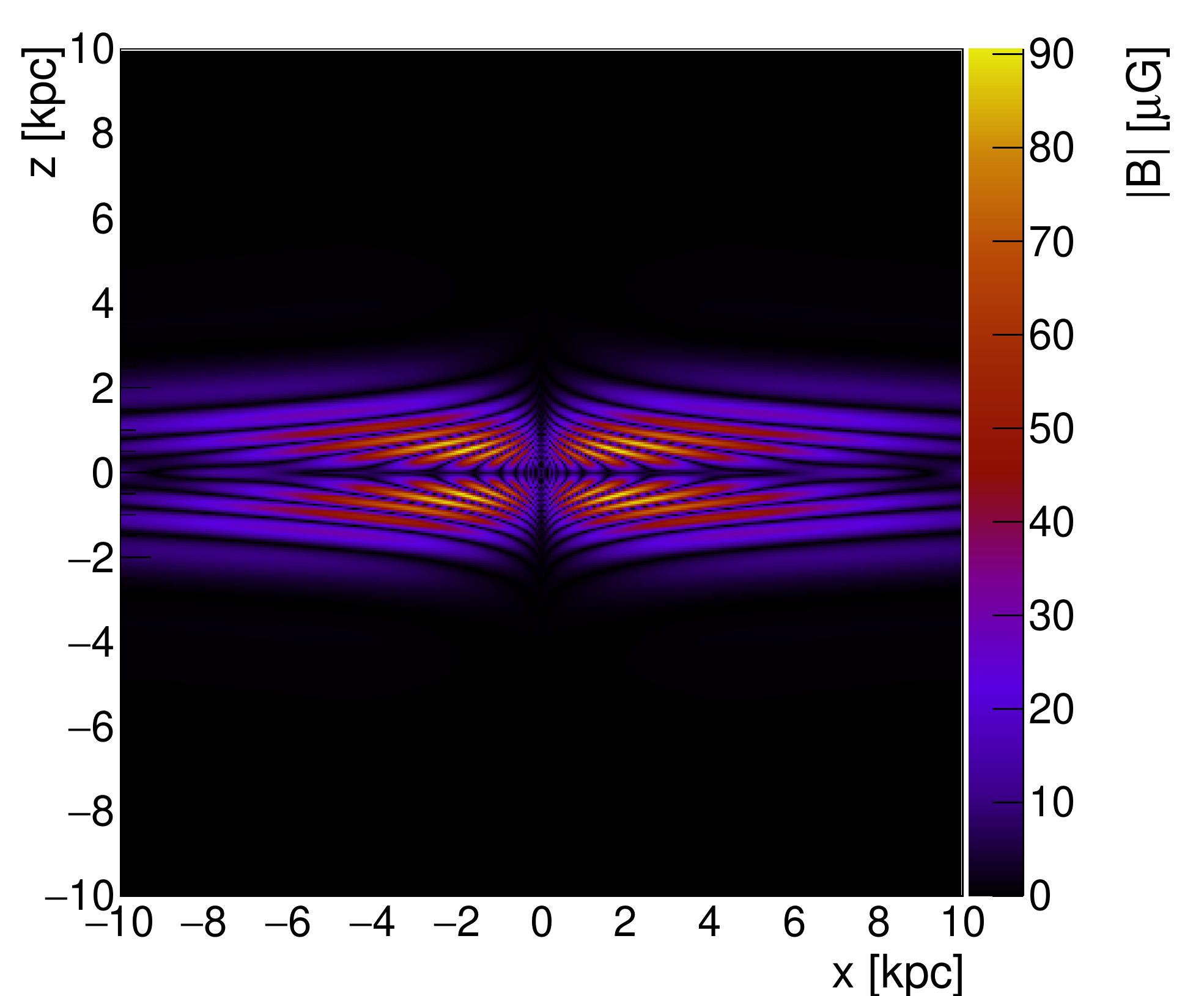}   
         
         \label{fig:Dd1C1XZ}    
     \end{subfigure}
        \caption{Same as Figure~\ref{fig:GMFJF12Planck} but for the Dd1C1 variant of the TF17 model of the GMF.}
    \label{fig:GMFDd1}
\end{figure}

The TF17 models of the GMF used in this work are combinations of three different models of the disk field (Ad1, Bd1, Dd1) together with a bisymmetric model of the halo field (C1) as suggested in \cite{TF17}. Visualizations of the three options of the TF17 field that fit the measured data the best in $xy$ plane at $z=0$~pc and in $xz$ plane at $y=0$~pc are shown in Figure~\ref{fig:GMFAd1}, Figure~\ref{fig:GMFBd1} and Figure~\ref{fig:GMFDd1}. Note that we show the field in the $xz$ plane only for a range from $-10$~kpc to $+10$~kpc as the field in this projection plane is mostly concentrated in this volume. Very strong amplitudes of the magnetic field appear locally in all presented models, therefore the results obtained with these models should be interpreted with caution. However, as our results show, the areas of possible extragalactic directions of the dipole are similar as the results obtained for the JF12Planck model of the GMF with the exception of the Bd1C1 option for the TF17 model. More detail about the individual options of the TF17 fields can be found in \cite{TF17}.

\begin{figure}
    \centering
    \includegraphics[width=\textwidth]{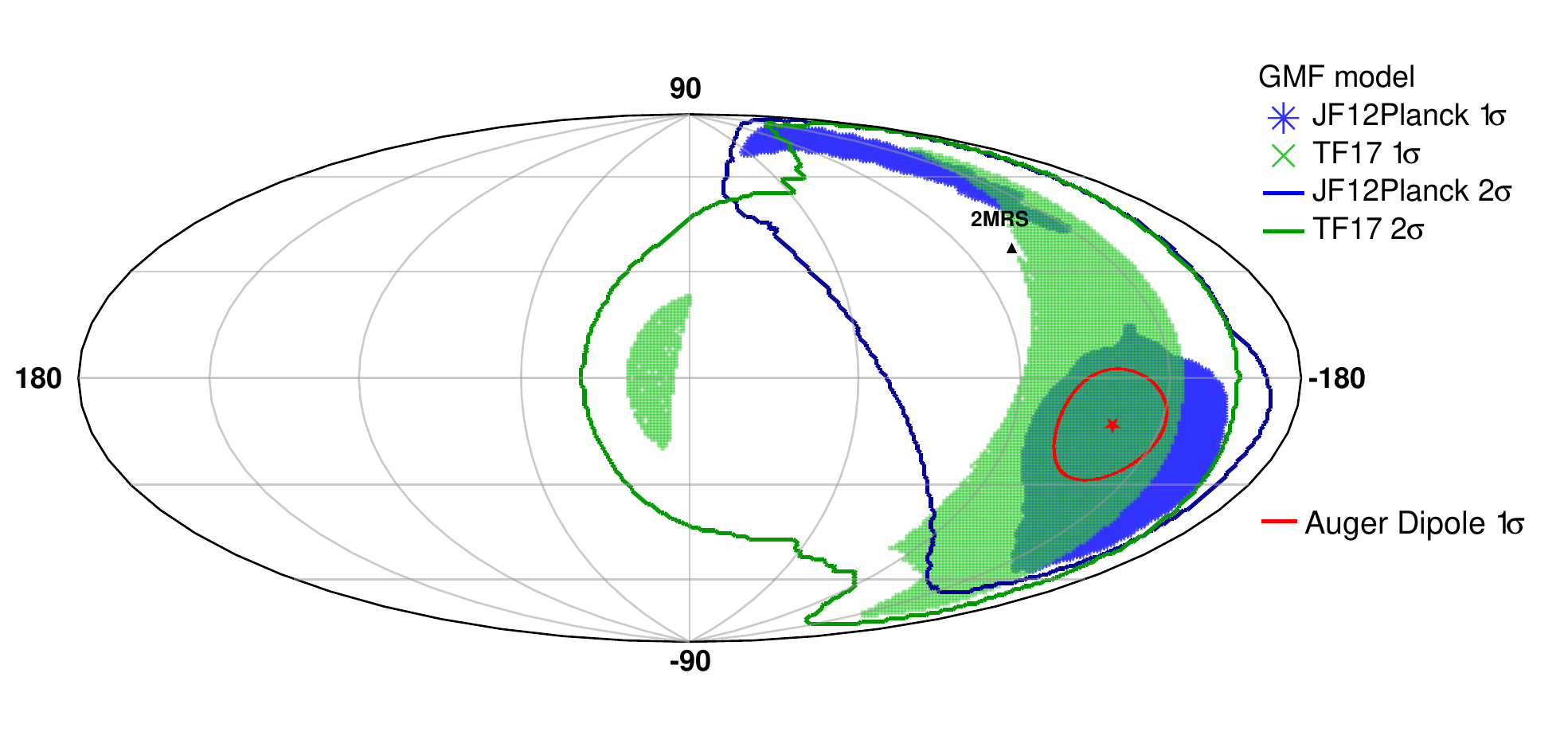}
    \caption{Same as Figure~\ref{fig:mixAll} with the strength of the three options of the TF17 model of the GMF adjusted by $\pm10\%$.}
    \label{fig:mixAlladjustedTF17}
\end{figure}

To account for uncertainties in the GMF models, we incorporate three distinct coherence lengths of the turbulent field in the case of the JF12Planck field. However, in the case of the three options of the TF17 model, the uncertainties were not taken into account in the previous results in Section~\ref{sec:results}. To assess how the results change under differing the field strength, we simulate particle propagation in the three options of the TF17 model (Ad1C1, Bd1C1 and Dd1C1) with the strength of the field adjusted by $\pm10\%$ for both the disk and halo components of the field. The combined results (as in Figure~\ref{fig:mixAll}) of the possible directions of the extragalactic dipole are shown in Figure~\ref{fig:mixAlladjustedTF17}. The area of the allowed extragalactic directions of the dipole by the TF17 model with adjustments by $\pm10\%$ is now slightly larger compared to Figure~\ref{fig:mixAll}. Nevertheless, no significantly different behavior is observed. In the case of the JF12Planck model of the GMF, no such modifications of the field strength were taken into account as the JF12 is multi-parameter model and not all of the  parameters are independent. Consequently, such ad-hoc changes can result in unexpected alterations, not only in the field strength but also in its overall structure and it can not be guaranteed that such modified field is still consistent with the observational values of Faraday rotations and polarized light that it was fitted to. An alternative method for addressing GMF uncertainties could involve employing the IMAGINE software package \cite{IMAGINE}.

\section{Extragalactic magnetic field}
\label{B:EGMF}

Deflections in the extragalactic magnetic field are not taken into account in this work as we are not trying to identify individual extragalactic sources and therefore the trajectory length of the cosmic rays from a source to our Galaxy is unknown. Moreover, these deflections are thought to be smaller than the deflections caused by the GMF. If we assume that the particles are coming from sources not further than 100 Mpc, the deflection in the extragalactic magnetic field can be estimated as \cite{EGdeflection_waxman, EGdeflection_Farrar} 
\begin{equation}
\delta\theta_{EG}\approx0.15^{\circ}\left(\frac{D}{3.8\rm{Mpc}}\cdot\frac{\lambda_{EG}}{100\rm{kpc}}  \right)^{\frac{1}{2}}\left(\frac{B_{EG}}{1\rm{nG}}\cdot\frac{Z}{E_{100}} \right),
\end{equation}
where $D$ is the distance of the source, $\lambda_{EG}$ and $B_{EG}$ are the coherence length and the strength of the extragalactic magnetic field, $Z$ is the charge of the cosmic ray and $E_{100}$ is the energy of the particle in units 100 EeV. This approximation of the total deflection is estimated for a particle experiencing many small deflections in a turbulent magnetic field. Taking a coherence length of the extragalactic magnetic field 100 kpc and strength of  1 nG, a proton of energy 100 EeV originating in a close source at $D=3.8$ Mpc would be deflected by $\approx0.15^{\circ}$, while a particle with the same energy originating in a source at $D=100$ Mpc would be deflected by $\approx 0.8^{\circ}$. Nevertheless, such particle would be also influenced by various energy-loss processes. 

\section{Spectral index}
\label{D:SI}

\begin{figure}
    \centering
    \includegraphics[width=1.0\textwidth]{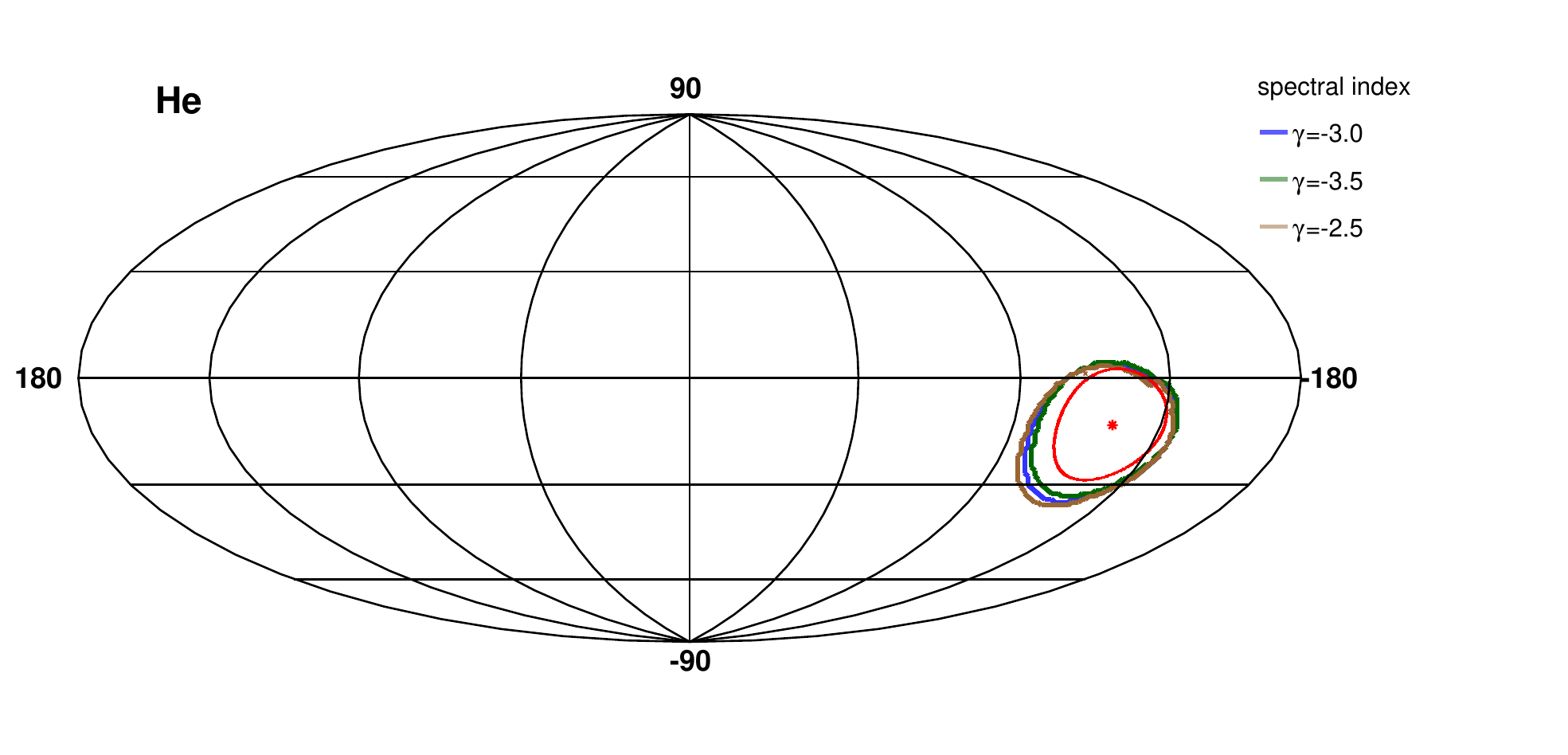}
         
    \caption{Allowed directions of the extragalactic dipole of cosmic rays above 8~EeV for allowed extragalactic directions at $1\sigma$ level in case of helium nuclei propagated in the JF12Planck model of GMF for three spectral indices. The red marker depicts the measured direction of the dipole on Earth together with the red $1\sigma$ contour.}
    \label{fig:spectralindex}
\end{figure}

Few simplifications are used in this work. One of these simplifications is the use of spectral index $\gamma=3$. This value was chosen as a close approximation of the spectral index of cosmic-ray energy spectrum above 8~EeV measured by current observatories  \cite{PAOenergyspectrum, TAenergyspectrum}. We verify that small changes of the spectral index do not significantly change our results. We confirm that by exploiting spectral indices 2.5 and 3.5 for pure proton and pure helium composition and one realisation of the JF12Planck model of the GMF. The resulting areas of allowed directions of the extragalactic dipole for lower and higher spectral indices differ by maximally $3^{\circ}$. The allowed directions of the extragalactic dipole for the three spectral indices and for the case of helium nuclei are shown in Figure~\ref{fig:spectralindex}. 

\section{Dipole and quadrupole in the right ascension}
\label{C:fit}

The allowed directions of the extragalactic dipole of particles above $8$~EeV were obtained using Equations \eqref{eq:reco3d} and compared with the measured direction and amplitude of the three-dimensional dipole according to \cite{AugerDipole2018} at the $1\sigma$ and $2\sigma$ level. In order to check the behavior of arrival directions in the right ascension, we also performed dipole and quadrupole fit of the distribution of arrival directions in the right ascension for all the found solutions.

\begin{figure}

      \begin{subfigure}{0.49\textwidth}
         \centering
          \includegraphics[width=1.0\textwidth]{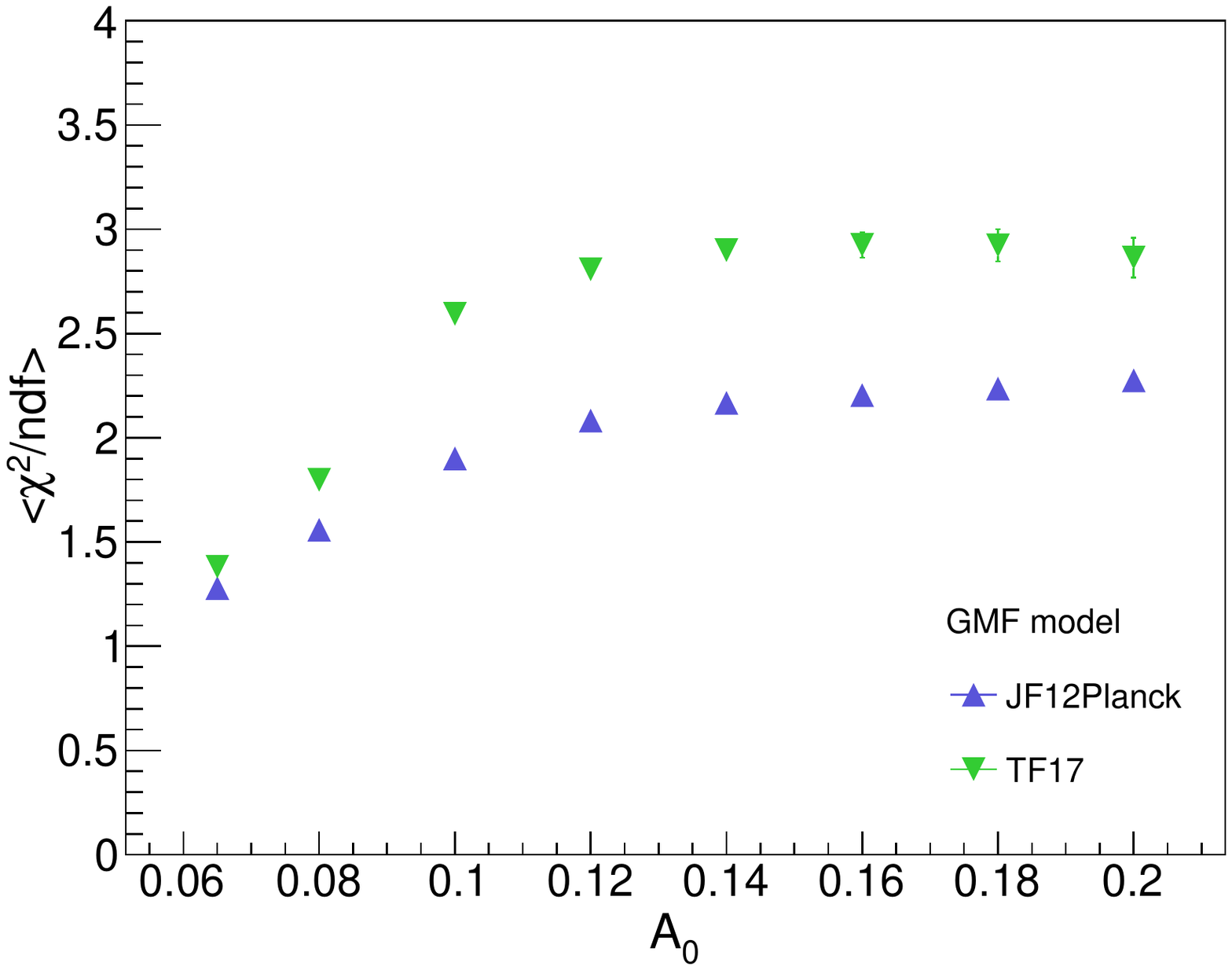}
         
         \label{fig:chi2vsA0}
     \end{subfigure}
     \hfill
     \begin{subfigure}{0.49\textwidth}
         \centering
         \includegraphics[width=1.0\textwidth]{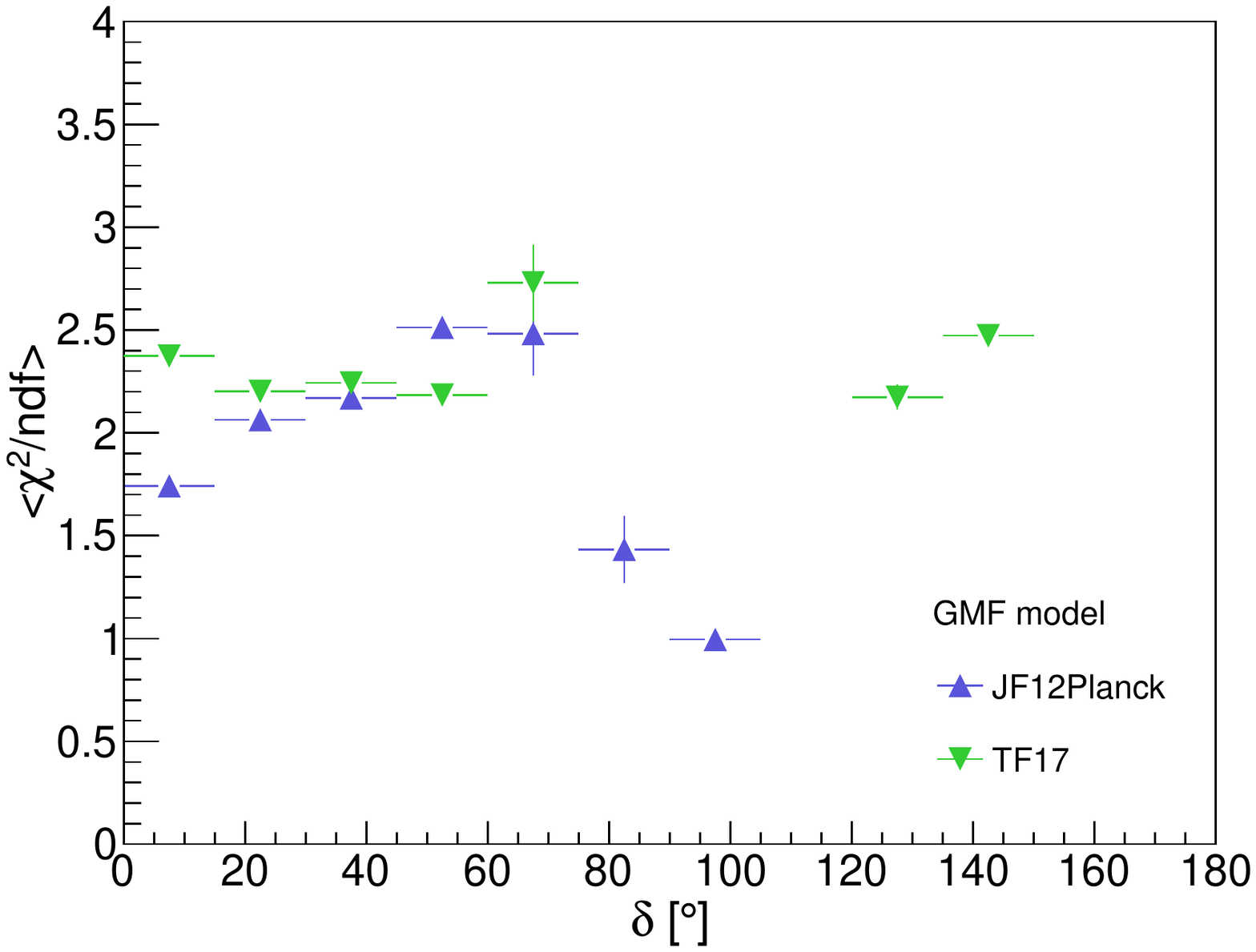}   
         
         \label{fig:chi2vsD}    
     \end{subfigure}
        \caption{The evolution of the mean $\chi^2/ndf$ of the dipole fit of the arrival directions of cosmic rays in the right ascension with initial amplitude of the dipole (left) and with the angular distance between the direction of the extragalactic dipole and the measured dipole direction on Earth. The results for the two models of the GMF are shown.}
    \label{fig:chi2}
\end{figure}

The distribution of the $\chi^2/ndf$ values for the dipole fit has a mean of $1.9$ with a standard deviation of $0.9$ for the JF12Planck model of the GMF, while the isotropy fit has a mean value of $\chi^2/ndf=72$ with a standard deviation of $24$. For the TF17 model of the GMF, the dipole fit has a mean value of $\chi^2/ndf=2.2$ with a standard deviation of $1.0$. The isotropy fit has a mean value of $\chi^2/ndf=79$ with a standard deviation $27$. The amplitude of the quadrupole is an order of magnitude lower than the dipole amplitude and is statistically insignificant (under $3\sigma$) for both models of the GMF in all studied cases. 

The evolution of the mean $\chi^2/ndf$ with the initial amplitude of the extragalactic dipole and with the angular distance between the direction of the extragalactic dipole and the measured dipole direction is shown in Figure~\ref{fig:chi2}.

\acknowledgments
This work was supported by the Ministry of Education, Youth and Sports of the Czech Republic – Grant No. LTT18004, LM2023032 and CZ.02.1.01/0.0/0.0/16\_013/0001402. We would like to thank our colleagues at FZU - Institute of Physics for numerous comments and suggestions and to Noémie Globus for useful suggestions regarding this work. We are also grateful for the stimulating environment within the Pierre Auger Collaboration.


\bibliographystyle{JHEP}
\bibliography{bibliography.bib}






\end{document}